\pdfoutput=1
\documentclass[11pt]{article}

\usepackage[preprint]{acl}
\usepackage{times}
\usepackage{latexsym}
\usepackage[T1]{fontenc}
\usepackage[utf8]{inputenc}
\usepackage{microtype}
\usepackage{inconsolata}
\usepackage{graphicx}

\usepackage{amsmath,amssymb,amsthm}
\usepackage{bm}
\usepackage{booktabs}
\usepackage{multirow}
\usepackage{makecell}
\usepackage{xcolor}
\usepackage{colortbl}
\usepackage{caption}
\usepackage{subcaption}
\usepackage{algorithm}
\usepackage{algpseudocode}
\usepackage[capitalise,noabbrev]{cleveref}

\usepackage{tikz}
\usetikzlibrary{arrows.meta,positioning,calc,fit,backgrounds,shapes.geometric,shadows.blur,decorations.pathreplacing}
\usepackage{pgfplots}
\usepackage{pgfplotstable}
\pgfplotsset{compat=1.18}
\usepgfplotslibrary{groupplots}

\definecolor{ours}{HTML}{C8102E}        % greedy selection under repair (RGS)
\definecolor{oursdark}{HTML}{6E0A1A}    % RGS with the task objective
\definecolor{ourslight}{HTML}{F4A3A8}   % repair Fisher score, no search (RFS)
\definecolor{replaceme}{HTML}{D9963A}
\definecolor{sleb}{HTML}{2A5D8F}
\definecolor{shortgpt}{HTML}{8A8A8A}
\definecolor{fullmodel}{HTML}{222222}
\definecolor{gridgrey}{HTML}{E6E6E6}
\definecolor{stream}{HTML}{3B6EA8}      % residual stream in the overview
\definecolor{bestcell}{HTML}{FBE3E6}
\definecolor{oursrow}{HTML}{FFF6F6}

\pgfplotsset{
  paperstyle/.style={
    width=\columnwidth, height=0.62\columnwidth,
    axis lines=left, axis line style={gray!70, line width=0.5pt},
    tick style={gray!70, line width=0.4pt},
    tick label style={font=\scriptsize}, label style={font=\footnotesize},
    legend style={font=\scriptsize, draw=none, fill=none},
    ymajorgrids, grid style={gridgrey, line width=0.4pt},
    every axis plot/.append style={line width=0.9pt},
  },
}

\newtheorem{proposition}{Proposition}
\DeclareMathOperator*{\argmin}{arg\,min}

\newcommand{\ours}{\textsc{Rgs}}
\newcommand{\oursfisher}{\textsc{Rfs}}

\newcommand{\best}[1]{\cellcolor{bestcell}\textbf{#1}}
\newcommand{\second}[1]{\underline{#1}}

\title{Repair Changes Which Layers to Remove:\\Exact Optima for Depth Pruning of Language Models}

\author{Jianchang Su \quad Dawei Xiang \quad Yifan Zhang \quad Wei Zhang \\
  University of Connecticut}

\begin{document}
\maketitle

\begin{abstract}
Depth pruning removes entire transformer layers to make large language models
faster, and recent methods recover quality by replacing the removed layers with
a training-free linear map. Which layers to remove, however, is chosen with
heuristic scores whose distance from the best choice is unknown. We evaluate
every subset of layer groups exactly (about 53,000 pruned models across four
families, with and without repair) and find that the best layers to remove
depend on the repair: linear repair absorbs the interaction among nearby layers
and turns the optimum from a split layout into a contiguous block, so methods
that choose layers without the repair miss the repaired optimum by up to 189\%.
We therefore select layers under the repair that will be applied. A greedy
search at the cost of SLEB lowers perplexity by 20\% to 66\% relative to
ReplaceMe on five models up to 27B parameters, and with a task objective it is
up to 25 points more accurate given the same data. A second-order expansion of
the loss yields a score that selects the repaired block from one forward and one
backward pass of the intact model.
\end{abstract}

\section{Introduction}
\label{sec:intro}

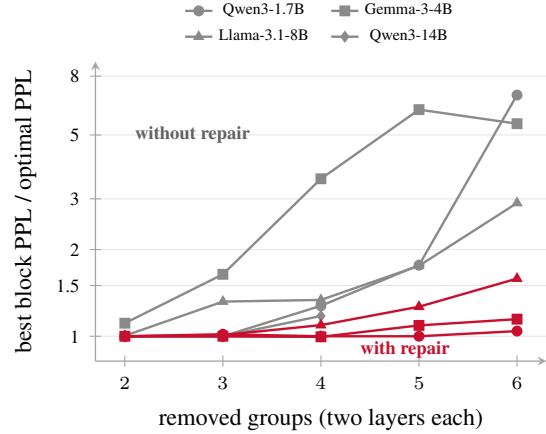
\begin{figure}[t]
\centering
% Teaser: perplexity of the best contiguous block relative to the exact optimum,
% without repair (grey) and with ReplaceMe's linear repair (red), for every
% exactly enumerated budget of the four models (one column).
% Data: data/teaser_block_excess.csv, written by data/export_csv.py.
\begin{tikzpicture}
\pgfplotstableread[col sep=comma,row sep=\\]{
model,model_index,removed_groups,removal_block_excess_pct,repair_block_excess_pct\\
Qwen3-1.7B,1,2,0.0,0.0\\
Qwen3-1.7B,1,3,0.0,1.6233094058823339\\
Qwen3-1.7B,1,4,27.593615616423172,0.0\\
Qwen3-1.7B,1,5,76.55084256850526,0.0\\
Qwen3-1.7B,1,6,588.011207358551,4.219379386563982\\
gemma-3-4b-it,2,2,11.006597992573774,0.0\\
gemma-3-4b-it,2,3,64.0899614934511,0.0\\
gemma-3-4b-it,2,4,252.3014484076251,-0.4439439246768373\\
gemma-3-4b-it,2,5,512.2937336369675,9.06264606566114\\
gemma-3-4b-it,2,6,447.1797611698318,14.683048343673821\\
Llama-3.1-8B-Instruct,3,2,0.4092671073749221,0.0\\
Llama-3.1-8B-Instruct,3,3,31.94923197456283,0.9763779045933729\\
Llama-3.1-8B-Instruct,3,4,33.727396409409,9.338913434701567\\
Llama-3.1-8B-Instruct,3,5,75.66228506998418,26.63031833391596\\
Llama-3.1-8B-Instruct,3,6,190.19385490201438,58.578755189771826\\
Qwen3-14B,4,2,0.0,0.0\\
Qwen3-14B,4,3,0.0,0.0\\
Qwen3-14B,4,4,17.637789374278245,0.0\\
}\teaser
\begin{axis}[
  paperstyle,
  width=0.98\columnwidth, height=0.72\columnwidth,
  ymode=log, ymin=0.82, ymax=9,
  ytick={1,1.5,2,3,5,8}, yticklabels={1,1.5,2,3,5,8}, yminorticks=false,
  ylabel={best block PPL / optimal PPL},
  xlabel={removed groups (two layers each)},
  xmin=1.7, xmax=6.3, xtick={2,3,4,5,6},
  legend style={at={(0.5,1.02)}, anchor=south, legend columns=2, font=\tiny,
    /tikz/every even column/.append style={column sep=6pt}},
  legend image code/.code={\draw[#1, mark repeat=2, mark phase=2] plot coordinates {(0cm,0cm) (0.18cm,0cm) (0.36cm,0cm)};},
  every axis plot/.append style={line width=0.9pt, mark size=1.6pt},
  clip=false,
]
% Without repair: one grey line per model.
\addplot[shortgpt, mark=*] table[x=removed_groups, y expr=1+\thisrow{removal_block_excess_pct}/100,
  restrict expr to domain={\thisrow{model_index}}{1:1}] {\teaser};
\addlegendentry{Qwen3-1.7B}
\addplot[shortgpt, mark=square*] table[x=removed_groups, y expr=1+\thisrow{removal_block_excess_pct}/100,
  restrict expr to domain={\thisrow{model_index}}{2:2}] {\teaser};
\addlegendentry{Gemma-3-4B}
\addplot[shortgpt, mark=triangle*] table[x=removed_groups, y expr=1+\thisrow{removal_block_excess_pct}/100,
  restrict expr to domain={\thisrow{model_index}}{3:3}] {\teaser};
\addlegendentry{Llama-3.1-8B}
\addplot[shortgpt, mark=diamond*] table[x=removed_groups, y expr=1+\thisrow{removal_block_excess_pct}/100,
  restrict expr to domain={\thisrow{model_index}}{4:4}] {\teaser};
\addlegendentry{Qwen3-14B}
% With repair: the same models in red.
\addplot[ours, mark=*, forget plot] table[x=removed_groups, y expr=1+\thisrow{repair_block_excess_pct}/100,
  restrict expr to domain={\thisrow{model_index}}{1:1}] {\teaser};
\addplot[ours, mark=square*, forget plot] table[x=removed_groups, y expr=1+\thisrow{repair_block_excess_pct}/100,
  restrict expr to domain={\thisrow{model_index}}{2:2}] {\teaser};
\addplot[ours, mark=triangle*, forget plot] table[x=removed_groups, y expr=1+\thisrow{repair_block_excess_pct}/100,
  restrict expr to domain={\thisrow{model_index}}{3:3}] {\teaser};
\addplot[ours, mark=diamond*, forget plot] table[x=removed_groups, y expr=1+\thisrow{repair_block_excess_pct}/100,
  restrict expr to domain={\thisrow{model_index}}{4:4}] {\teaser};
\node[font=\scriptsize\bfseries, text=shortgpt!70!black, anchor=west] at (axis cs:2.0,5.0) {without repair};
\node[font=\scriptsize\bfseries, text=ours, anchor=west] at (axis cs:4.3,0.9) {with repair};
\end{axis}
\end{tikzpicture}
\caption{Repair changes which set of removed layers is best. Perplexity of
the best contiguous block relative to the exact optimum over all subsets, at
every enumerated budget of four models. Without repair (grey) a block is up to
6.9 times worse than the optimum. With ReplaceMe's linear repair (red) the best
block is the optimum at 9 of the 18 budgets and within 5\% of it at 13.}
\label{fig:teaser}
\end{figure}

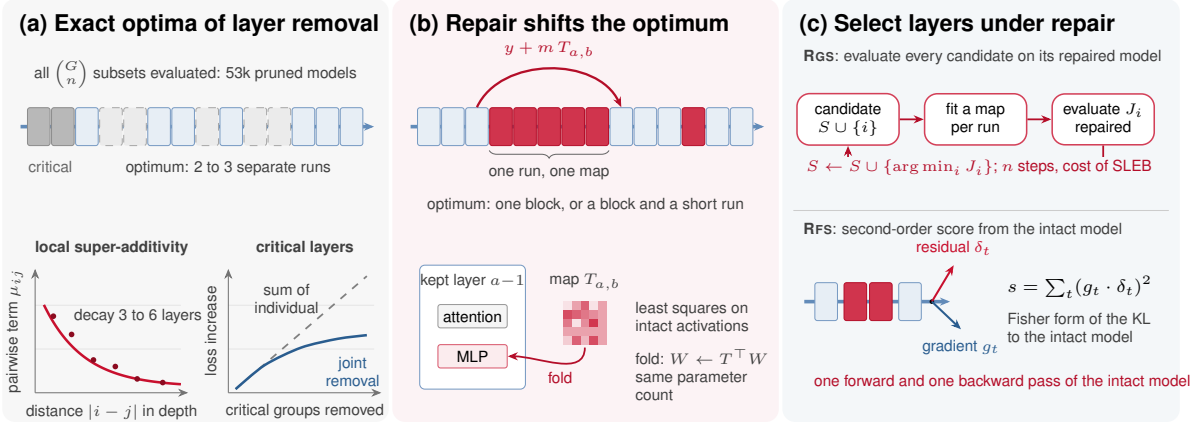
\begin{figure*}[t]
\centering
\resizebox{\textwidth}{!}{% System overview: (a) exact optima reveal the interaction structure,
% (b) linear repair reverses it, (c) selection under repair.
\begin{tikzpicture}[
  font=\sffamily\scriptsize,
  >={Stealth[length=4pt,width=3.2pt]},
  layer/.style={rounded corners=1pt, minimum width=3.1mm, minimum height=5.2mm,
                inner sep=0pt, draw=black!45, line width=0.35pt, fill=white},
  kept/.style={layer, fill=stream!12, draw=stream!70},
  cut/.style={layer, fill=black!8, draw=black!35, dashed},
  cutours/.style={layer, fill=ours!85, draw=ours!90!black},
  sat/.style={layer, fill=black!28, draw=black!45},
  panel/.style={rounded corners=5pt, fill=#1, draw=none},
  title/.style={font=\sffamily\bfseries\small, anchor=north west},
  note/.style={font=\sffamily\tiny, text=black!75, align=center},
  stage/.style={draw=ours!80, fill=white, rounded corners=5pt, line width=0.6pt,
               minimum width=1.4cm, minimum height=0.66cm, align=center, font=\sffamily\tiny},
  stream line/.style={line width=1.4pt, stream!80, ->},
  axis/.style={black!55, line width=0.4pt, ->},
]

% ---------------------------------------------------------------- panel (a)
\fill[panel=black!3] (-0.25,-4.05) rectangle (5.05,1.75);
\node[title] at (-0.15,1.68) {(a) Exact optima of layer removal};

\draw[stream line] (0,0) -- (4.85,0);
\foreach \i/\s in {0/sat,1/sat,2/kept,3/cut,4/cut,5/kept,6/kept,7/cut,8/kept,9/cut,10/cut,11/kept,12/kept,13/kept} {
  \node[\s] (a\i) at ({0.25+0.33*\i},0) {};
}
\node[note, anchor=south] at (2.4,0.4) {all $\binom{G}{n}$ subsets evaluated: 53k pruned models};
\node[note, anchor=north, text=black!60] at (0.41,-0.33) {critical};
\node[note, anchor=north] at (2.85,-0.33) {optimum: 2 to 3 separate runs};

% Inset: local super-additivity decays with distance.
\begin{scope}[shift={(0.2,-3.65)}]
  \draw[gridgrey] (0,0.6) -- (2.0,0.6) (0,1.2) -- (2.0,1.2);
  \draw[axis] (0,0) -- (2.1,0);
  \draw[axis] (0,0) -- (0,1.7);
  \node[note, anchor=north] at (1.05,-0.02) {distance $|i-j|$ in depth};
  \node[note, rotate=90, anchor=south] at (-0.03,0.85) {pairwise term $\mu_{ij}$};
  \draw[ours, line width=1pt, smooth, domain=0.12:2.0, samples=30] plot (\x, {1.4*exp(-\x/0.55)+0.08});
  \foreach \x/\y in {0.25/1.05,0.5/0.8,0.8/0.45,1.1/0.36,1.4/0.19,1.75/0.14} {
    \fill[ours!70!black] (\x,\y) circle (1.1pt);
  }
  \node[note, anchor=south, font=\sffamily\tiny\bfseries] at (1.05,1.75) {local super-additivity};
  \node[note, anchor=west] at (0.4,1.05) {decay 3 to 6 layers};
\end{scope}

% Inset: critical layers are sub-additive.
\begin{scope}[shift={(2.85,-3.65)}]
  \draw[gridgrey] (0,0.6) -- (2.0,0.6) (0,1.2) -- (2.0,1.2);
  \draw[axis] (0,0) -- (2.1,0);
  \draw[axis] (0,0) -- (0,1.7);
  \node[note, anchor=north] at (1.05,-0.02) {critical groups removed};
  \node[note, rotate=90, anchor=south] at (-0.03,0.85) {loss increase};
  \draw[black!50, dashed, line width=0.7pt] (0.1,0.05) -- (1.9,1.6);
  \node[note, anchor=south, font=\sffamily\tiny\bfseries] at (1.05,1.75) {critical layers};
  \node[note] at (0.8,1.3) {sum of\\individual};
  \draw[sleb, line width=1pt, smooth] plot coordinates {(0.1,0.05) (0.5,0.38) (1.0,0.62) (1.5,0.74) (1.9,0.79)};
  \node[note, text=sleb, anchor=north west] at (1.2,0.62) {joint\\removal};
\end{scope}

% ---------------------------------------------------------------- panel (b)
\begin{scope}[shift={(5.35,0)}]
  \fill[panel=ours!5] (-0.25,-4.05) rectangle (5.05,1.75);
  \node[title] at (-0.15,1.68) {(b) Repair shifts the optimum};

  \draw[stream line] (0,0) -- (4.85,0);
  \foreach \i/\s in {0/kept,1/kept,2/kept,3/cutours,4/cutours,5/cutours,6/cutours,7/cutours,8/kept,9/kept,10/kept,11/cutours,12/kept,13/kept} {
    \node[\s] (b\i) at ({0.25+0.33*\i},0) {};
  }
  % The run is bypassed through the map T of layer a-1.
  \draw[ours!90!black, line width=0.9pt, ->] (b2.north) to[out=65, in=115] node[pos=0.5, above, note, text=ours!90!black] {$y + m\,T_{a,b}$} (b8.north);
  \draw[decorate, decoration={brace, amplitude=3pt, mirror}, black!60]
    ($(b3.south west)+(0,-0.06)$) -- ($(b7.south east)+(0,-0.06)$)
    node[midway, below=3pt, note] {one run, one map};
  \node[note, anchor=north] at (2.4,-0.85) {optimum: one block, or a block and a short run};

  % What the repair does inside kept layer a-1.
  \begin{scope}[shift={(0.15,-3.75)}]
    \node[draw=stream!70, fill=white, rounded corners=2pt, minimum width=1.45cm, minimum height=1.7cm,
          line width=0.4pt] (lay) at (0.7,1.0) {};
    \node[note, anchor=north] at (lay.north) {kept layer $a{-}1$};
    \node[draw=black!40, fill=black!4, rounded corners=1pt, minimum width=0.95cm, minimum height=0.32cm,
          inner sep=0pt, font=\sffamily\tiny] (attn) at (0.7,1.15) {attention};
    \node[draw=ours!70, fill=ours!8, rounded corners=1pt, minimum width=0.95cm, minimum height=0.32cm,
          inner sep=0pt, font=\sffamily\tiny] (mlp) at (0.7,0.6) {MLP};
    % The map T as a small matrix.
    \foreach \r in {0,...,4} \foreach \c in {0,...,4} {
      \pgfmathsetmacro{\shade}{int(mod(7*\r+3*\c+5*\r*\c,9))*9+12}
      \fill[ours!\shade] ({1.95+0.12*\c},{1.35-0.12*\r}) rectangle ++(0.11,-0.11);
    }
    \node[note, anchor=south] at (2.24,1.37) {map $T_{a,b}$};
    \draw[->, ours!90!black, line width=0.7pt] (2.24,0.72) to[out=-90, in=0] node[pos=0.65, below right, note, text=ours!90!black, xshift=-2pt] {fold} (mlp.east);
    \node[note, anchor=west, align=left] at (2.8,1.15)
      {least squares on\\intact activations};
    \node[note, anchor=west, align=left] at (2.8,0.35)
      {fold: $W \gets T^{\top} W$\\same parameter\\count};
  \end{scope}
\end{scope}

% ---------------------------------------------------------------- panel (c)
\begin{scope}[shift={(10.7,0)}]
  \fill[panel=sleb!5] (-0.25,-4.05) rectangle (5.2,1.75);
  \node[title] at (-0.15,1.68) {(c) Select layers under repair};

  \node[note, anchor=west, align=left] at (-0.1,0.95) {\textbf{\textsc{Rgs}}: evaluate every candidate on its repaired model};
  \node[stage] (cand) at (0.65,0.1) {candidate\\$S \cup \{i\}$};
  \node[stage] (fit) at (2.4,0.1) {fit a map\\per run};
  \node[stage] (eval) at (4.15,0.1) {evaluate $J_i$\\repaired};
  \draw[->, ours!90!black, line width=0.7pt] (cand) -- (fit);
  \draw[->, ours!90!black, line width=0.7pt] (fit) -- (eval);
  \draw[->, ours!90!black, line width=0.7pt, rounded corners=6pt]
    (eval.south) -- ++(0,-0.4) -| (cand.south);
  \node[note, text=ours!90!black, fill=sleb!5, inner sep=1pt] at (2.45,-0.56)
    {$S \gets S \cup \{\arg\min_i J_i\}$; $n$ steps, cost of SLEB};

  % RFS: second-order score from the intact model.
  \draw[black!20, line width=0.4pt] (-0.1,-1.15) -- (5.05,-1.15);
  \node[note, anchor=west, align=left] at (-0.1,-1.4) {\textbf{\textsc{Rfs}}: second-order score from the intact model};
  \begin{scope}[shift={(0.05,-3.2)}]
    \draw[stream line] (0,0.8) -- (2.0,0.8);
    \node[kept] at (0.3,0.8) {};
    \node[cutours] at (0.7,0.8) {};
    \node[cutours] at (1.05,0.8) {};
    \node[kept] at (1.45,0.8) {};
    \fill[black] (1.75,0.8) circle (0.9pt);
    \draw[->, ours, line width=0.8pt] (1.75,0.8) -- ++(0.3,0.5) node[note, anchor=south, text=ours] {residual $\delta_t$};
    \draw[->, sleb, line width=0.8pt] (1.75,0.8) -- ++(0.4,-0.38) node[note, anchor=north, text=sleb] {gradient $g_t$};
    \node[anchor=west, font=\sffamily\scriptsize] at (2.65,1.0) {$s = \sum_t (g_t \cdot \delta_t)^2$};
    \node[note, anchor=west, align=left] at (2.65,0.45)
      {Fisher form of the KL\\to the intact model};
    \node[note, anchor=west, align=left, text=ours!90!black] at (0,-0.3)
      {one forward and one backward pass of the intact model};
  \end{scope}
\end{scope}

\end{tikzpicture}}
\caption{Overview. (a) Evaluating every subset of layer groups shows critical
layers and local super-additivity, so the optimum without repair splits the
removal into separate runs. (b) ReplaceMe's least-squares map replaces a run by
a linear map folded into the down-projection of the kept layer before it; it
absorbs the interaction inside the run, and the repaired optimum becomes one
block. (c) \ours{} evaluates every candidate on its repaired model at the cost of
SLEB; \oursfisher{} ranks blocks by the Fisher-weighted residual of their repair,
computed on the intact model only.}
\label{fig:overview}
\end{figure*}

Depth pruning removes whole transformer layers. Unlike width pruning or
sparsity, it reduces latency and memory on standard hardware without special
kernels, because the pruned model is a shorter model of the same architecture
\citep{men2025shortgpt,song2024sleb,kim2024shortened}. Recent methods add a
training-free repair: they replace the removed layers by a linear map fitted on
calibration activations and fold the map into the weights of a kept layer, so
the repaired model has the same parameter count as the pruned one
\citep{shopkhoev2025replaceme,chen2025simple,chen2026prune}.

Every depth-pruning method must choose which layers to remove. For $L$ layers
and $n$ removed layers there are $\binom{L}{n}$ choices, and each choice must be
scored on a pruned model. Published methods therefore use proxies: the
similarity of a layer's input and output
\citep{men2025shortgpt,gromov2025unreasonable,shopkhoev2025replaceme}, greedy
removal \citep{song2024sleb}, evolutionary search \citep{sieberling2025evopress},
or game-theoretic estimates of each layer's contribution
\citep{siddiqui2024deeper,ding2026pruning}. How far these choices are from the
best one is unknown, and recent work concludes that more elaborate search adds
little \citep{kim2026rethinking}.

These proxies score each layer in the unrepaired model. With a linear repair,
the interactions among the removed layers change: a map absorbs the joint effect
of contiguous layers, so the layers that are best to remove together differ from
those individually least important. The gap between the proxy and the true
optimum therefore depends on the repair, and it is unknown because the true
optimum has not been computed.

We close this gap. At a granularity of two-layer groups the number of subsets
is small enough to evaluate every one: about 53,000 pruned models over
Qwen3-1.7B, Gemma-3-4B, Llama-3.1-8B and Qwen3-14B, each scored after plain
removal and after the least-squares repair of ReplaceMe
\citep{shopkhoev2025replaceme}. The exact optima show that the answer depends on
the repair (\cref{fig:teaser}): without repair, removing two nearby layers costs
more than the sum of their individual costs, so the best subset spreads the
removal over separate runs and a contiguous block is close to the worst choice.
With repair, one map absorbs the interaction inside a run, and the best subset
becomes a block, or a block and one short run. Published methods choose layers
without the repair, and so miss the repaired optimum by up to 189\%.

The fix follows directly: evaluate every candidate subset on the repaired model
that will be deployed (\cref{fig:overview}). A greedy search that does so costs
as much as SLEB and finds the exact repaired optimum, or comes within 0.2\% of
it, at every enumerated budget on three of four models. A second-order expansion
of the repaired loss turns the same principle into a score computed from one
forward and one backward pass of the intact model. Our contributions are:

\begin{itemize}
\item \textbf{The structure of the exact optimum.} A Möbius decomposition of
  the loss increase from removal shows two interactions in all four families: a
  few \emph{critical} layers whose joint loss increase is far below the sum of
  their individual loss increases, and \emph{local super-additivity} among all other layers, which decays
  with distance in depth with a length of 3 to 6 layers (\cref{sec:structure}).
  As a result, at 4 to 6 removed groups the best contiguous block is 18\% to
  588\% worse than the optimum.
\item \textbf{Linear repair makes the optimum contiguous.} One least-squares map per run
  absorbs the interaction among the run's layers, and the repaired optimum
  becomes one block, or one block and one short run (\cref{sec:reversal}).
  Selection must therefore account for the repair: choosing layers without the
  repair and then repairing them is 5\% to 33\% worse than the repaired optimum, and
  ReplaceMe's cosine criterion is 11\% to 189\% worse.
\item \textbf{Selection methods.} Greedy selection under repair (\ours{}), at
  the cost of SLEB, has 20\% to 66\% lower perplexity than ReplaceMe on four
  models and 54\% to 56\% lower on a 27B hybrid-attention model, and with a
  task objective is more accurate at every budget, by up to 25 points on the
  27B model. A second-order expansion of
  the Kullback--Leibler divergence gives a score (\oursfisher{}) that selects
  the repaired block from the gradients of the intact model
  (\cref{sec:method}).
\item \textbf{Evaluation beyond the selection data.} We test the methods with
  ReplaceMe's default calibration, on tasks that the objective does not see, on
  C4, after LoRA recovery, and on a 27B model that was not used for any design
  choice, and we report the cases where a baseline is better
  (\cref{sec:experiments}).
\end{itemize}

\section{Related work}
\label{sec:related}

\paragraph{Choosing layers to remove.}
ShortGPT removes the layers whose output is most similar to their input
\citep{men2025shortgpt}; \citet{gromov2025unreasonable} remove the contiguous
block with the smallest angular distance between its input and output, and
LLM-Streamline \citep{chen2025streamlining}, LaCo \citep{yang2024laco} and CLP
\citep{lu2025structural} remove or merge contiguous blocks by design. SLEB removes
layers greedily by the perplexity of the pruned model \citep{song2024sleb};
BlockPruner does the same for attention and MLP sublayers
\citep{zhong2025blockpruner}. EvoPress searches over subsets and observes that
removal is not additive \citep{sieberling2025evopress}, and Shapley-value
estimates model the interaction with sampling or surrogates
\citep{siddiqui2024deeper,ding2026pruning}. \citet{kim2026rethinking} call
the subset problem NP-hard and find that search gains little over one-shot
scores without repair. We compute the optimum exactly, decompose its
interactions, and show that the gain of search depends on the repair.

\paragraph{Training-free repair.}
ReplaceMe fits one linear map per removed block by least squares and folds it
into the down-projection of the last kept layer \citep{shopkhoev2025replaceme}.
LinearPatch \citep{chen2025simple} and Prune\&Comp
\citep{chen2026prune} correct scale and outliers at the pruning interface;
Prune\&Comp alternates pruning and compensation, so each candidate is scored on a
model that earlier steps have compensated, with one scalar per layer. Other
repairs use low-rank or shifted maps
\citep{liu2025grasp,yun2026ghosted,bahri2026shift}, and SubFit selects
submodules by the residual of their fitted replacement
\citep{cunegatti2026from}. These methods repair after the layers are chosen,
or choose with a criterion that does not evaluate the repaired model. ReplaceMe
reports that its cosine criterion finds optimal or near-optimal 8-layer blocks
on Llama~3~8B \citep[App.~A.10]{shopkhoev2025replaceme}; over all block lengths
and four families we find it 11\% to 189\% worse in perplexity than the
repaired optimum.

\paragraph{Second-order saliency and model reduction.}
Optimal Brain Damage and Optimal Brain Surgeon score the removal of a weight by a
second-order expansion of the loss \citep{lecun1990optimal,hassibi1993second}, and OBC,
GPTQ and SparseGPT apply it layer by layer
\citep{frantar2022optimal,frantar2023optq,frantar2023sparsegpt}. Fisher pruning and
Taylor importance score the removal of an activation $a$ by $\sum (g \cdot a)^2$
\citep{theis2018faster,molchanov2019importance}, and BRECQ weights
reconstruction error by the Fisher information \citep{li2021brecq}. Our score
applies this expansion to the error that remains after repair. Balanced
truncation reduces a linear system by weighting the error of the reduced model by
how much the output observes it \citep{moore1981principal}; our score has the
same form, with the loss gradient in the role of the observability.

\section{Problem setup}
\label{sec:setup}

\paragraph{Pruning and repair.}
A decoder-only model with $L$ layers updates a residual stream $h_\ell \in
\mathbb{R}^d$ at each token by $h_\ell = h_{\ell-1} + f_\ell(h_{\ell-1})$. We
partition the layers into $G$ contiguous groups of one or two layers. Removing a
set $S$ of groups skips their layers; a \emph{run} is a maximal set of adjacent
removed groups. Linear repair \citep{shopkhoev2025replaceme} replaces a run of
layers $a..b$ by a map $T \in \mathbb{R}^{d \times d}$ on the MLP output
$m_{a-1}$ of the last kept layer before the run, which then outputs $y +
m_{a-1}T$ instead of $y + m_{a-1}$, with $y = h_{a-1} - m_{a-1}$. The map solves
\begin{equation}
\label{eq:repair}
T_{a,b} = \argmin_T \big\| M_{a-1} T - (H_b - Y_{a-1}) \big\|_F^2 + \lambda \|T\|_F^2,
\end{equation}
where the rows of $M_{a-1}$, $Y_{a-1}$ and $H_b$ are activations of the intact
model at calibration tokens. $T$ folds into the down-projection of layer $a-1$,
so the repaired model has the architecture of the pruned one. Each run has its
own map; a run that starts at the first layer cannot be repaired.

\paragraph{Objective and data.}
The \emph{loss increase} of a subset is the mean token loss of the pruned model
minus that of the intact model on 16 WikiText-2 calibration sequences of 256 tokens \citep{merity2017pointer}. Every method
selects layers on these sequences, and we report perplexity (PPL) on 16 other
sequences of the WikiText-2 test split. Maps are fitted on 128 WikiText-2 training
sequences with $\lambda = 10^{-3}$ times the mean diagonal of $M^\top M$.
Accuracy is the mean over ARC-Easy, ARC-Challenge, HellaSwag, PIQA and
WinoGrande \citep{clark2018think,zellers2019hellaswag,bisk2020piqa,sakaguchi2020winogrande},
1,000 items each, with the prompts of lm-evaluation-harness
\citep{sutawika2023eleutherai}; one standard error of the mean is about 0.7 points.
BoolQ \citep{clark2019boolq} is reported separately, because pruning shifts the
yes/no prior and moves it by 20 points or more independently of the other tasks.

\paragraph{Exact enumeration.}
We use Qwen3-1.7B and Qwen3-14B \citep{qwenteam2025qwen3}, Gemma-3-4B
\citep{gemmateam2025gemma} and Llama-3.1-8B-Instruct
\citep{llamateam2024llama} with two-layer groups (14, 17, 16 and 20 groups),
and evaluate every subset of up to 6 removed groups (4 on Qwen3-14B), without and
with repair (\cref{tab:enumeration}). We also enumerate Qwen3-1.7B at
single-layer granularity (28 groups) up to 3 removed layers; the same
interaction structure appears at both granularities. We store the loss of every
subset on every sequence, so any search method can be evaluated against the
exact optimum without further model runs.

\section{The structure of the exact optimum}
\label{sec:structure}

\paragraph{Interaction decomposition.}
Let $c(S)$ be the loss increase from removing $S$, with $c(\emptyset) = 0$. Its Möbius
transform \citep{grabisch1999axiomatic}
\begin{equation}
\label{eq:mobius}
\mu(S) = \sum_{R \subseteq S} (-1)^{|S|-|R|} c(R), \quad c(S) = \sum_{R \subseteq S} \mu(R)
\end{equation}
splits the loss increase into an individual term $\mu(\{i\})$ for each group and an interaction
term for each larger set; removal is additive exactly when every interaction
term is zero. The \emph{interaction share} of a budget, the mean absolute
difference between $c(S)$ and $\sum_{i \in S} c(\{i\})$ as a share of the mean
absolute loss increase, rises from 21\% to 50\% at 2 to 6 removed groups on Qwen3-1.7B,
15\% to 31\% on Gemma, 7\% to 25\% on Llama and 11\% to 22\% (2 to 4 removed)
on Qwen3-14B.

\paragraph{Critical layers.}
We call a group \emph{critical} when its removal alone more than doubles
perplexity: 4 of 14 groups in Qwen3-1.7B, 5 of 17 in Gemma, 3 of 16 in Llama
and 3 of 20 in Qwen3-14B, usually the first ones. Their pairwise terms are large
and negative, because the loss saturates once the model fails, and every exact
optimum keeps them. We call the other groups \emph{non-critical}.

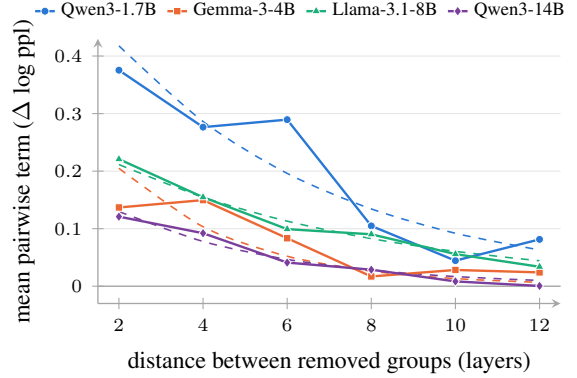
\begin{figure}[t]
\centering
% Mean pairwise interaction term among non-critical groups against their distance
% in layers, one line per model, with the exponential fit over the first four
% distances dashed (one column).
% Data: data/interaction_decay.csv, written by data/export_csv.py.
\begin{tikzpicture}
% One colour per model; the series of this figure are models, not methods.
\definecolor{mqwensmall}{HTML}{2A78D6}
\definecolor{mgemma}{HTML}{EB6834}
\definecolor{mllama}{HTML}{1BAF7A}
\definecolor{mqwenlarge}{HTML}{7A3E9D}
\pgfplotstableread[col sep=comma,row sep=\\]{
model,model_index,distance,mean_pair_term,fitted,amplitude,decay_length\\
Qwen3-1.7B,1,2,0.37539750151336193,0.41771757131075943,0.6097173285758,5.2883482940054165\\
Qwen3-1.7B,1,4,0.2764531203678676,0.2861784653379211,0.6097173285758,5.2883482940054165\\
Qwen3-1.7B,1,6,0.289465069770813,0.1960609743233422,0.6097173285758,5.2883482940054165\\
Qwen3-1.7B,1,8,0.10479659587144852,0.13432144730816215,0.6097173285758,5.2883482940054165\\
Qwen3-1.7B,1,10,0.044469013810157776,0.09202367411070922,0.6097173285758,5.2883482940054165\\
Qwen3-1.7B,1,12,0.08141078799962997,0.06304545377184474,0.6097173285758,5.2883482940054165\\
gemma-3-4b-it,2,2,0.13685847683386368,0.20490688734737503,0.40631655225769087,2.9215126570520487\\
gemma-3-4b-it,2,4,0.1496474027633667,0.10333527455155528,0.40631655225769087,2.9215126570520487\\
gemma-3-4b-it,2,6,0.08339769144852956,0.05211234773452382,0.40631655225769087,2.9215126570520487\\
gemma-3-4b-it,2,8,0.016977883875370026,0.026280442938669844,0.40631655225769087,2.9215126570520487\\
gemma-3-4b-it,2,10,0.028256369488579885,0.013253321162407085,0.40631655225769087,2.9215126570520487\\
gemma-3-4b-it,2,12,0.024038222928841908,0.0066836971600448165,0.40631655225769087,2.9215126570520487\\
Llama-3.1-8B-Instruct,3,2,0.22080236052473387,0.21141902674614302,0.28908768532933304,6.392064050562376\\
Llama-3.1-8B-Instruct,3,4,0.15467357635498047,0.15461746431490397,0.28908768532933304,6.392064050562376\\
Llama-3.1-8B-Instruct,3,6,0.09919252246618271,0.11307667355727594,0.28908768532933304,6.392064050562376\\
Llama-3.1-8B-Instruct,3,8,0.09023278620507982,0.08269657091735295,0.28908768532933304,6.392064050562376\\
Llama-3.1-8B-Instruct,3,10,0.055820949375629425,0.0604786347736416,0.28908768532933304,6.392064050562376\\
Llama-3.1-8B-Instruct,3,12,0.03379737266472408,0.044229950813063384,0.28908768532933304,6.392064050562376\\
Qwen3-14B,4,2,0.12091961205005645,0.12978623649864518,0.21653481436350555,3.9072987322497115\\
Qwen3-14B,4,4,0.0921351718051093,0.07779103436089858,0.21653481436350555,3.9072987322497115\\
Qwen3-14B,4,6,0.041099290762628825,0.04662624628152828,0.21653481436350555,3.9072987322497115\\
Qwen3-14B,4,8,0.028731842453663167,0.027946753249478415,0.21653481436350555,3.9072987322497115\\
Qwen3-14B,4,10,0.008368554214636484,0.01675067326825849,0.21653481436350555,3.9072987322497115\\
Qwen3-14B,4,12,0.0005126094276254827,0.0100399875590266,0.21653481436350555,3.9072987322497115\\
}\decay
\begin{axis}[
  paperstyle,
  height=0.66\columnwidth,
  ymin=-0.03, ymax=0.43,
  ytick={0,0.1,0.2,0.3,0.4},
  yticklabel style={/pgf/number format/fixed, /pgf/number format/precision=1},
  ylabel={mean pairwise term ($\Delta$ log ppl)},
  xlabel={distance between removed groups (layers)},
  xmin=1.5, xmax=12.5,
  xtick={2,4,6,8,10,12},
  legend style={at={(0.44,1.03)}, anchor=south, legend columns=-1,
    inner xsep=0pt, nodes={inner xsep=1.5pt},
    /tikz/every even column/.append style={column sep=2pt},
    /tikz/every odd column/.append style={column sep=0.5pt}},
  legend image code/.code={
    \draw[#1, mark=none, solid] (0cm,0cm) -- (0.26cm,0cm);
    \draw[#1, only marks] plot coordinates {(0.13cm,0cm)};},
]
\draw[gray!70, line width=0.5pt] (axis cs:1.5,0) -- (axis cs:12.5,0);
\pgfplotsinvokeforeach{1,2,3,4}{
  \addplot[forget plot, smooth, dashed, line width=0.6pt,
    color={\ifcase#1\or mqwensmall\or mgemma\or mllama\or mqwenlarge\fi},
    restrict expr to domain={\thisrow{model_index}}{#1:#1}]
    table[x=distance, y=fitted] {\decay};
}
\addplot[mqwensmall, mark=*, mark size=1.5pt, mark options={draw=white, line width=0.3pt},
  restrict expr to domain={\thisrow{model_index}}{1:1}]
  table[x=distance, y=mean_pair_term] {\decay};
\addlegendentry{Qwen3-1.7B}
\addplot[mgemma, mark=square*, mark size=1.4pt, mark options={draw=white, line width=0.3pt},
  restrict expr to domain={\thisrow{model_index}}{2:2}]
  table[x=distance, y=mean_pair_term] {\decay};
\addlegendentry{Gemma-3-4B}
\addplot[mllama, mark=triangle*, mark size=1.9pt, mark options={draw=white, line width=0.3pt},
  restrict expr to domain={\thisrow{model_index}}{3:3}]
  table[x=distance, y=mean_pair_term] {\decay};
\addlegendentry{Llama-3.1-8B}
\addplot[mqwenlarge, mark=diamond*, mark size=1.9pt, mark options={draw=white, line width=0.3pt},
  restrict expr to domain={\thisrow{model_index}}{4:4}]
  table[x=distance, y=mean_pair_term] {\decay};
\addlegendentry{Qwen3-14B}
\end{axis}
\end{tikzpicture}
\caption{Mean pairwise interaction term of two removed non-critical groups
against their distance in depth, with exponential fits over the first four
distances (dashed). Removing two nearby groups costs more than the sum of their
individual costs.}
\label{fig:interaction}
\end{figure}

\paragraph{Local super-additivity.}
Among non-critical groups the pairwise term $\mu(\{i,j\})$ is positive and decays with
the distance $|i-j|$ in depth (\cref{fig:interaction}), with fitted decay lengths
of 2.9 layers (Gemma) to 6.4 layers (Llama) and terms close to zero beyond about
10 layers. Pairwise terms explain 52\% to 76\% of the non-additivity at 3
removed groups; higher-order terms among nearby groups grow at heavier budgets.

\paragraph{A contiguous block is close to the worst choice.}
A block contains the most pairs within the interaction range, so under local
super-additivity it has the largest interaction cost. At small budgets the
optimum can be a block; from 4 removed groups on, every exact optimum in every
family splits the removal into 2 or 3 runs, and the best block is 18\% to 588\%
worse in perplexity (\cref{tab:block}). Methods that remove one block by
design \citep{gromov2025unreasonable,chen2025streamlining,lu2025structural} therefore
cannot find the optimum when the layers are not repaired.

\begin{table}[t]
\centering
\small
\begin{tabular}{lrrrr}
\toprule
 & \multicolumn{4}{c}{Removed groups} \\
\cmidrule(l){2-5}
Model & 3 & 4 & 5 & 6 \\
\midrule
Qwen3-1.7B & 0\% & 28\% & 77\% & 588\% \\
Gemma-3-4B & 64\% & 252\% & 512\% & 447\% \\
Llama-3.1-8B & 32\% & 34\% & 76\% & 190\% \\
Qwen3-14B & 0\% & 18\% & -- & -- \\
\bottomrule
\end{tabular}
\caption{Excess perplexity of the best contiguous block over the exact
optimum without repair.}
\label{tab:block}
\end{table}

\paragraph{Implications for existing methods.}
One-shot scoring removes the groups with the smallest individual loss increase.
These cluster in depth, so one-shot scoring removes an almost contiguous set:
33\% above the optimum on Qwen3-1.7B and 581\% on Gemma at 4 removed groups. A
pairwise model assigns large negative terms to critical-layer pairs; its optimum
is 189 to 226,000 times worse. SLEB evaluates actual pruned models and accounts
for the interaction, but cannot undo an earlier removal: it finds the optimum in
13 of 20 budgets (\cref{app:search}). The interaction remains when a removed
layer is replaced by its mean update; it therefore comes from the lost
computation (\cref{app:mechanism}).

\section{Linear repair makes the optimum contiguous}
\label{sec:reversal}

\Cref{tab:exact-repair} compares the exact optimum without repair, that optimum
after repair, ReplaceMe's choice, and the exact optimum under repair over every
subset of non-critical groups.

\begin{table}[t]
\centering
\small
\setlength{\tabcolsep}{3.5pt}
\resizebox{\columnwidth}{!}{%
\begin{tabular}{@{}lrrrrr@{}}
\toprule
 & & \multicolumn{2}{c}{Removal opt.} & & Repaired \\
\cmidrule(lr){3-4}
Model & $n$ & as is & repaired & ReplaceMe & opt. \\
\midrule
Qwen3-1.7B & 4 & 73.4 & 50.3 & 75.1 & \best{41.7} \\
\quad (26.5) & 5 & 128.7 & 59.2 & 90.6 & \best{54.3} \\
 & 6 & 229.2 & 80.2 & 99.5 & \best{76.0} \\
\midrule
Gemma-3-4B & 4 & 46.7 & 50.6 & 44.7 & \best{38.3} \\
\quad (32.0) & 5 & 65.9 & 51.6 & 72.2 & \best{42.8} \\
 & 6 & 93.5 & 64.1 & 149.8 & \best{51.9} \\
\midrule
Llama-3.1-8B & 4 & 33.6 & 26.8 & 40.4 & \best{22.2} \\
\quad (13.7) & 5 & 53.0 & 32.5 & 61.9 & \best{28.3} \\
 & 6 & 84.2 & 44.1 & 94.0 & \best{35.1} \\
\midrule
Qwen3-14B (15.0) & 4 & 23.1 & 25.6 & 21.3 & \best{19.1} \\
\bottomrule
\end{tabular}}
\caption{Perplexity at the heaviest enumerated budgets ($n$ removed
groups; full model in parentheses). ``Removal opt., repaired'' repairs the best
subset for plain removal. The last column is the exact optimum under repair.}
\label{tab:exact-repair}
\end{table}

\paragraph{Repair absorbs the interaction inside a run.}
One map is fitted to the input and output of a whole run, so the
super-additivity among the run's layers is part of what it fits. At the heaviest
budget the interaction share falls from 62\% to 36\% on Qwen3-1.7B, 66\% to
52\% on Gemma, 57\% to 28\% on Llama and 25\% to 15\% on Qwen3-14B. Separately
repaired runs still interact: the sum of their repaired loss increases explains
the loss increase of a subset with $R^2$ of 0.8 to 0.9 at 2 removed groups but only $-0.09$ to
0.44 at 6, because each map assumes that the rest of the network is intact.

\paragraph{The repaired optimum is contiguous.}
Without repair, every optimum at 4 or more removed groups splits into 2 or 3
runs. Under repair it is one block (Qwen3-1.7B at 4 and 5 removed groups,
Qwen3-14B at 4) or one block and one more run (\cref{fig:layout} in
\cref{app:tables}). A contiguous block, which is close to the worst choice
without repair, is close to the best choice with repair.

\paragraph{Selection must account for the repair.}
Two simpler approaches are worse. Repairing the best subset for plain removal is 5\% to 33\%
worse than the repaired optimum, because that subset is split into runs that
repair cannot absorb. Choosing the block with the smallest cosine distance
between its input and output, as ReplaceMe does, is 11\% to 189\% worse. On
Llama at 8 of 32 layers, a setting that ReplaceMe evaluates, its block is
among the most accurate subsets under the WikiText objective (\cref{tab:main}),
in agreement with \citet[App.~A.10]{shopkhoev2025replaceme}, but its perplexity
is 79\% higher than that of the subset chosen under repair, and with a task
objective the subset chosen under repair is 4.0 points more accurate. A
selection method must therefore evaluate each subset on the repaired model that
it deploys.

\section{Selecting layers under repair}
\label{sec:method}

\subsection{Greedy selection under repair}
\label{sec:greedy}

We therefore include the repair in the search. Greedy selection under repair
(\ours{}) removes one layer at a time, as SLEB does, but evaluates each
candidate subset after fitting and applying its repair maps (\cref{alg:rgs}).
Maps are fitted once per run from cached activations of the intact model, so a
candidate costs one forward pass over the calibration text and \ours{} costs as
many evaluations as SLEB, $\sum_{k=0}^{n-1}(L-k)$ for $n$ removed layers. On the
exact tables \ours{} finds the repaired optimum at every budget on Qwen3-1.7B
and Qwen3-14B, and is within 0.2\% of it on Llama and within 3.4\% on Gemma. The objective can be any loss on the repaired model.
With a \emph{task objective}, the cross-entropy of the correct choice among the
choices on 200 training items per task (\ours{}-T), it selects for accuracy.

\subsection{A second-order score from the intact model}
\label{sec:score}

Under repair the optimum is close to one block, so ranking blocks is
sufficient, and a second-order expansion of the loss ranks them with the
activations and gradients of the intact model. Let $z \in \mathbb{R}^{N \times
d}$ be the residual stream after
block $a..b$ at the $N$ calibration tokens of the intact model. The repaired
model produces $z + \delta$ there, with $\delta_t = y_t + m_t T_{a,b} - z_t$ the
residual of the repair at token $t$; the layers after the block are unchanged.

\begin{proposition}
\label{prop:kl}
Let $p_s(z)$ be the next-token distribution at position $s$ as a function of the
residual stream, and $D(\delta) = \sum_s \mathrm{KL}(p_s(z) \,\|\, p_s(z+\delta))$. Then
$D(0)=0$, $\nabla D(0) = 0$ and
\begin{equation}
\label{eq:kl}
D(\delta) = \tfrac12 \operatorname{vec}(\delta)^\top F \operatorname{vec}(\delta) + O(\|\delta\|^3),
\end{equation}
where $F = \sum_s \mathbb{E}_{y \sim p_s}[\nabla_z \log p_s(y)\, \nabla_z \log p_s(y)^\top]$
is the Fisher information of the downstream network with respect to the residual stream.
\end{proposition}

The proof is in \cref{app:proof}. We replace the expectation over $y$ by the
observed tokens (the empirical Fisher) and keep one rank-one block per position,
$F \approx \operatorname{blockdiag}_t(g_t g_t^\top)$ with $g_t = \nabla_{z_t}
\sum_s \ell_s$ the gradient of the summed token loss, so that each position
plays the role of a sample as in Fisher pruning \citep{theis2018faster} and
Taylor importance \citep{molchanov2019importance}. The result is the
\emph{repair Fisher score} (\oursfisher{}) of a block:
\begin{equation}
\label{eq:score}
s(a,b) = \sum_{t=1}^{N} (g_t \cdot \delta_t)^2.
\end{equation}
The least-squares map of \cref{eq:repair} minimises the Euclidean norm of
$\delta$; \cref{eq:score} measures the same residual in the metric of the
downstream Fisher information, which weights each direction of the residual stream
by how much the loss responds to it. This is the weighting that balanced truncation
applies to the error of a reduced system, with the observability Gramian in the
role of $F$ \citep{moore1981principal}. The unweighted residual and the
first-order term $\sum_t g_t \cdot \delta_t$ give inconsistent block rankings
across models (\cref{sec:exp-score}).

\section{Experiments}
\label{sec:experiments}

\subsection{Experimental setup}
\label{sec:exp-setup}

\paragraph{Models and budgets.}
We use the four enumerated families of \cref{sec:setup} and Qwen3.8-27B, which
has 64 layers with three linear-attention layers to each full-attention layer
and was not used for any design choice. We remove 25\% and 37.5\% of the layers
at single-layer granularity.

\paragraph{Baselines.}
ShortGPT's block influence \citep{men2025shortgpt}, LaCo
\citep{yang2024laco}, ReplaceMe's cosine block \citep{shopkhoev2025replaceme},
Prune\&Comp \citep{chen2026prune} and SLEB \citep{song2024sleb}. Every
method is repaired with the same least-squares maps (+R), so the methods differ
only in which layers they choose. For LaCo we remove the layers that its merge
criterion selects, because this is better than its parameter merge at every
budget we test. For the task objective we also give ReplaceMe's criterion the
same task training texts.

\paragraph{Metrics and implementation.}
We report WikiText-2 perplexity on the test sequences, C4 perplexity, and the
average accuracy over five zero-shot tasks (\cref{sec:setup}). All models run in
bfloat16 on one 48\,GB GPU, except Qwen3.8-27B, which runs on two
(\cref{app:details}).

\begin{table*}[t!]
\centering
\setlength{\tabcolsep}{2.6pt}
\resizebox{\textwidth}{!}{%
\begin{tabular}{l rrrr c rrrr c rrrr c rrrr c rrrr}
\toprule
 & \multicolumn{4}{c}{Qwen3-1.7B} & & \multicolumn{4}{c}{Gemma-3-4B} & & \multicolumn{4}{c}{Llama-3.1-8B} & & \multicolumn{4}{c}{Qwen3-14B} & & \multicolumn{4}{c}{Qwen3.8-27B} \\
\cmidrule(lr){2-5}\cmidrule(lr){7-10}\cmidrule(lr){12-15}\cmidrule(lr){17-20}\cmidrule(l){22-25}
Removed & \multicolumn{2}{c}{7/28} & \multicolumn{2}{c}{10/28} & & \multicolumn{2}{c}{8/34} & \multicolumn{2}{c}{13/34} & & \multicolumn{2}{c}{8/32} & \multicolumn{2}{c}{12/32} & & \multicolumn{2}{c}{10/40} & \multicolumn{2}{c}{15/40} & & \multicolumn{2}{c}{16/64} & \multicolumn{2}{c}{24/64} \\
Method & PPL$\downarrow$ & Avg$\uparrow$ & PPL$\downarrow$ & Avg$\uparrow$ & & PPL$\downarrow$ & Avg$\uparrow$ & PPL$\downarrow$ & Avg$\uparrow$ & & PPL$\downarrow$ & Avg$\uparrow$ & PPL$\downarrow$ & Avg$\uparrow$ & & PPL$\downarrow$ & Avg$\uparrow$ & PPL$\downarrow$ & Avg$\uparrow$ & & PPL$\downarrow$ & Avg$\uparrow$ & PPL$\downarrow$ & Avg$\uparrow$ \\
\midrule
Full model & 26.5 & 60.0 & 26.5 & 60.0 & & 32.0 & 69.8 & 32.0 & 69.8 & & 13.7 & 72.4 & 13.7 & 72.4 & & 15.0 & 73.0 & 15.0 & 73.0 & & 10.1 & 72.8 & 10.1 & 72.8 \\
\midrule
\multicolumn{25}{l}{\textit{Same calibration text for every method (WikiText-2)}} \\
ShortGPT + R & 77.4 & \second{46.1} & 130.0 & 40.4 & & 44.7 & 50.8 & 87.2 & \second{42.7} & & 39.4 & \best{59.5} & 94.0 & 46.8 & & 48.2 & 49.2 & 266.4 & 37.7 & & 34.8 & 40.6 & 42.4 & 38.6 \\
LaCo + R & 78.7 & 39.9 & 571.3 & 36.2 & & 57.0 & \second{54.0} & 445.4 & 39.6 & & 46.7 & \second{58.5} & 82.6 & 45.6 & & 90.9 & 41.1 & 1119.1 & 36.4 & & n/a & n/a & n/a & n/a \\
ReplaceMe & 53.7 & 45.6 & 68.7 & 40.9 & & 44.7 & 50.8 & 150.7 & 41.5 & & 40.4 & 57.2 & 94.0 & 46.8 & & 41.9 & 51.4 & 43.3 & 44.5 & & 28.7 & 42.2 & 36.9 & 42.2 \\
Prune\&Comp + R & 63.5 & \best{46.7} & 120.2 & \second{41.2} & & 44.7 & 50.8 & 192.3 & 40.6 & & 42.2 & 57.0 & 87.0 & 45.6 & & 45.4 & 51.4 & 76.6 & 42.2 & & 28.1 & 48.9 & 33.9 & 42.9 \\
SLEB + R & 40.3 & 45.9 & \second{59.1} & \best{41.8} & & 50.5 & \best{54.7} & \second{69.1} & \best{44.7} & & \second{23.6} & 56.2 & \second{37.8} & \second{48.4} & & \second{27.2} & 49.1 & 43.1 & 43.0 & & \best{13.2} & \second{65.7} & 17.1 & \second{57.4} \\
\rowcolor{oursrow} \oursfisher{} (ours) & \second{38.7} & 44.9 & \best{54.3} & 41.1 & & \second{38.2} & 48.1 & 97.6 & 40.6 & & 24.3 & 51.2 & 95.9 & 39.4 & & \best{21.9} & \second{53.0} & \best{34.0} & \best{45.8} & & 13.4 & \best{66.4} & \best{15.7} & 55.5 \\
\rowcolor{oursrow} \ours{} (ours) & \best{37.7} & 44.2 & \best{54.3} & 41.1 & & \best{35.9} & 50.6 & \best{51.1} & \best{44.7} & & \best{22.6} & 55.4 & \best{34.4} & \best{49.0} & & \best{21.9} & \best{54.4} & \second{34.1} & \second{45.1} & & \second{13.3} & 64.9 & \second{16.0} & \best{57.9} \\
\textit{\ours{} vs.\ ReplaceMe} & \textit{$-$30\%} & \textit{$-$1.5} & \textit{$-$21\%} & \textit{+0.2} & & \textit{$-$20\%} & \textit{$-$0.2} & \textit{$-$66\%} & \textit{+3.2} & & \textit{$-$44\%} & \textit{$-$1.8} & \textit{$-$63\%} & \textit{+2.2} & & \textit{$-$48\%} & \textit{+3.0} & \textit{$-$21\%} & \textit{+0.6} & & \textit{$-$54\%} & \textit{+22.7} & \textit{$-$56\%} & \textit{+15.7} \\
\midrule
\multicolumn{25}{l}{\textit{Selection on task training data$^\dagger$}} \\
ReplaceMe, task data & \best{51.9} & 43.8 & \best{68.7} & 40.9 & & 71.3 & 45.8 & 192.3 & 40.6 & & 40.9 & 58.9 & 94.0 & 46.8 & & 41.9 & 51.4 & 251.3 & 38.8 & & 28.7 & 42.2 & 36.9 & 42.2 \\
\rowcolor{oursrow} \ours{}-T (ours) & 64.5 & \best{47.1} & 110.0 & \best{41.7} & & \best{68.1} & \best{59.9} & \best{143.7} & \best{51.2} & & \best{34.6} & \best{61.2} & \best{54.0} & \best{51.9} & & \best{37.9} & \best{59.5} & \best{67.7} & \best{47.9} & & \best{27.6} & \best{66.8} & \best{33.3} & \best{59.8} \\
\textit{\ours{}-T vs.\ ReplaceMe, task data} & \textit{+24\%} & \textit{+3.3} & \textit{+60\%} & \textit{+0.9} & & \textit{$-$5\%} & \textit{+14.1} & \textit{$-$25\%} & \textit{+10.6} & & \textit{$-$15\%} & \textit{+2.3} & \textit{$-$43\%} & \textit{+5.1} & & \textit{$-$10\%} & \textit{+8.1} & \textit{$-$73\%} & \textit{+9.1} & & \textit{$-$4\%} & \textit{+24.6} & \textit{$-$10\%} & \textit{+17.6} \\
\bottomrule
\end{tabular}}
\caption{Main results: WikiText-2 perplexity (PPL) and the average zero-shot
accuracy (\%) over ARC-Easy, ARC-Challenge, HellaSwag, PIQA and WinoGrande
(Avg; one standard error is about 0.7 points). Every method is repaired with the
same least-squares maps (+R); the methods differ only in which layers they
remove. The upper block selects on WikiText-2 calibration text; the lower block
($^\dagger$) selects on the training items of the five tasks, which are disjoint
from the evaluation items. Within each block the best result is highlighted and
the second best underlined; italic rows give the change relative to ReplaceMe.
LaCo's merge needs identical layers and does not apply to the hybrid 27B model.
Per-task scores are in \cref{tab:per-task}.}
\label{tab:main}
\end{table*}

\subsection{Main results}
\label{sec:exp-main}

\paragraph{Perplexity.}
On the four enumerated families \ours{} has the lowest perplexity at 7
of 8 settings, with \oursfisher{} 0.3\% closer at the eighth; it is 20\% to 66\%
below ReplaceMe and 4\% to 29\% below SLEB with repair
(\cref{tab:main}; \cref{fig:main-ppl} in \cref{app:tables}). SLEB uses the same
greedy search without repair, so this difference comes from the repair in the
search. It is 20\% to 73\% below Prune\&Comp and 37\% to 97\% below
LaCo, both with repair, on every model they apply to. The gap to ReplaceMe is largest where the repaired optimum is
not one block: on Gemma at 13 of 34 layers \ours{} removes a block of 8 layers
and 5 separate ones and has one third of ReplaceMe's perplexity. On Qwen3.8-27B
it is 54\% and 56\% below ReplaceMe, whose criterion picks the first layers on
this architecture; \ours{} matches SLEB with repair at 16 layers (within 0.5\%)
and is 7\% better at 24. Across all five models and both budgets, including the
repair in the search consistently lowers perplexity.

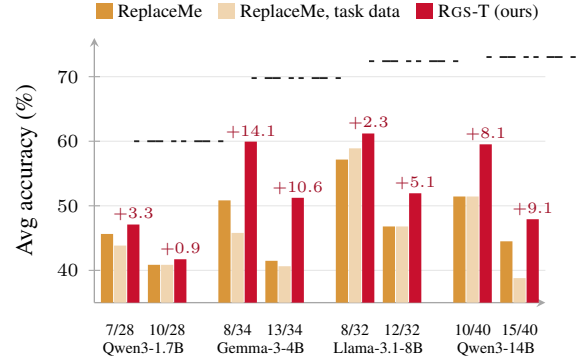
\begin{figure}[t]
\centering
% Five-task mean accuracy (%) of ReplaceMe, ReplaceMe given the task data, and
% RGS-T, as grouped bars; the dashed line marks the full model and the label
% above each RGS-T bar is its gain over ReplaceMe given the same data (one column).
% Data: data/main_accuracy.csv, written by data/export_csv.py.
\begin{tikzpicture}
\pgfplotstableread[col sep=comma,row sep=\\]{
x,model,label,removed,depth,budget,full,replaceme,replaceme_task_data,greedy_repair,greedy_tasks,span_low,span_width,tasks_gain_points,tasks_gain_over_task_data_points,full_pct,replaceme_pct,replaceme_task_data_pct,greedy_tasks_pct\\
1.0,Qwen3-1.7B,Qwen3-1.7B,7,28,7/28,0.6,0.45640000000000003,0.4384,0.4418,0.471,0.4418,0.029199999999999948,1.4599999999999946,3.2599999999999962,60.0,45.64,43.84,47.099999999999994\\
2.0,Qwen3-1.7B,Qwen3-1.7B,10,28,10/28,0.6,0.40859999999999996,0.40859999999999996,0.4109999999999999,0.41720000000000007,0.40859999999999996,0.008600000000000108,0.8600000000000108,0.8600000000000108,60.0,40.86,40.86,41.720000000000006\\
3.5,gemma-3-4b-it,Gemma-3-4B,8,34,8/34,0.6978,0.5084,0.458,0.5064,0.5993999999999999,0.5064,0.09299999999999997,9.099999999999998,14.139999999999992,69.78,50.839999999999996,45.800000000000004,59.93999999999999\\
4.5,gemma-3-4b-it,Gemma-3-4B,13,34,13/34,0.6978,0.41479999999999995,0.4064,0.4466,0.5124000000000001,0.41479999999999995,0.09760000000000013,9.760000000000012,10.600000000000009,69.78,41.48,40.64,51.24000000000001\\
6.0,Llama-3.1-8B-Instruct,Llama-3.1-8B,8,32,8/32,0.724,0.5716,0.589,0.5536,0.612,0.5536,0.05840000000000001,4.039999999999999,2.300000000000002,72.39999999999999,57.16,58.9,61.199999999999996\\
7.0,Llama-3.1-8B-Instruct,Llama-3.1-8B,12,32,12/32,0.724,0.46799999999999997,0.46799999999999997,0.49000000000000005,0.5194,0.46799999999999997,0.0514,5.140000000000001,5.140000000000001,72.39999999999999,46.8,46.8,51.94\\
8.5,Qwen3-14B,Qwen3-14B,10,40,10/40,0.7303999999999999,0.5144,0.5144,0.5439999999999999,0.5952,0.5144,0.08079999999999998,8.079999999999998,8.079999999999998,73.03999999999999,51.44,51.44,59.519999999999996\\
9.5,Qwen3-14B,Qwen3-14B,15,40,15/40,0.7303999999999999,0.445,0.388,0.4514,0.47919999999999996,0.445,0.03419999999999995,3.4199999999999955,9.119999999999994,73.03999999999999,44.5,38.800000000000004,47.919999999999995\\
}\mainacc
\begin{axis}[
  paperstyle,
  width=0.98\columnwidth, height=0.66\columnwidth,
  ybar, bar width=0.26,
  ymin=35, ymax=76,
  ytick={40,50,60,70},
  ylabel={Avg accuracy (\%)},
  xmin=0.45, xmax=10.05,
  xtick={1,2,3.5,4.5,6,7,8.5,9.5},
  xticklabels from table={\mainacc}{budget},
  x tick label style={font=\tiny},
  major x tick style={draw=none},
  extra x ticks={1.5,4,6.5,9},
  extra x tick labels={Qwen3-1.7B,Gemma-3-4B,Llama-3.1-8B,Qwen3-14B},
  extra x tick style={major tick length=0pt, tick label style={font=\scriptsize, yshift=-11pt}},
  extra x tick style={grid=none},
  legend style={at={(0.5,1.02)}, anchor=south, legend columns=-1,
    /tikz/every even column/.append style={column sep=4pt}},
  legend image code/.code={\draw[#1, draw=none] (0cm,-0.08cm) rectangle (0.22cm,0.12cm);},
  clip=false,
]
\addplot[bar shift=-0.28, fill=replaceme, draw=none]
  table[x=x, y=replaceme_pct] {\mainacc};
\addlegendentry{ReplaceMe}
\addplot[bar shift=0, fill=replaceme!40, draw=none]
  table[x=x, y=replaceme_task_data_pct] {\mainacc};
\addlegendentry{ReplaceMe, task data}
\addplot[bar shift=0.28, fill=ours, draw=none,
  nodes near coords={\pgfmathprintnumber[showpos, fixed, precision=1]{\pgfplotspointmeta}},
  nodes near coords style={font=\tiny, text=ours!80!black, anchor=south, inner sep=1pt},
  point meta=explicit]
  table[x=x, y=greedy_tasks_pct, meta=tasks_gain_over_task_data_points] {\mainacc};
\addlegendentry{\ours{}-T (ours)}
% Full model, one dashed segment per point.
\addplot[draw=none, forget plot, mark=none,
  error bars/.cd, x dir=both, x fixed=0.45, error mark=none,
  error bar style={fullmodel, dashed, line width=0.7pt}]
  table[x=x, y=full_pct] {\mainacc};
\end{axis}
\end{tikzpicture}
\caption{Average accuracy over the five zero-shot tasks. Labels give the gain of
\ours{}-T over ReplaceMe's criterion given the same task data (0.9 to 14.1
points); dashed lines mark the full model.}
\label{fig:main-acc}
\end{figure}

\paragraph{Accuracy.}
Given the same task training data, \ours{}-T is more accurate than ReplaceMe's
criterion at all 8 settings, by 0.9 to 14.1 points (lower block of
\cref{tab:main}, \cref{fig:main-acc}), and also has lower perplexity at 6 of
them. Both methods use the same data, so the gain comes from selection under
repair. \ours{}-T has the highest average of all methods at 7 of 8 settings and
beats plain ReplaceMe at all 8, by 0.9 to 9.8 points; on Gemma at 8 of 34
layers it keeps 86\% of the full model's accuracy against ReplaceMe's 73\%.

The accuracy gain transfers to the 27B model and to tasks outside the
objective. On Qwen3.8-27B, \ours{}-T keeps 92\% of the full model's accuracy at
16 of 64 layers and 82\% at 24, 24.6 and 17.6 points above ReplaceMe given the
same data, and above every other method at both budgets. WikiText perplexity and
task accuracy favour different layers (\cref{tab:per-task}). Built from ARC-Easy,
HellaSwag and PIQA only, the task objective still beats ReplaceMe on
ARC-Challenge and WinoGrande at both budgets on three of four models
(\cref{tab:unseen}). Selection under repair thus improves both perplexity and
accuracy, and the gain generalises across models, budgets and tasks.

\paragraph{MMLU.}
The selection objective decides which layers are kept. 5-shot MMLU
\citep{hendrycks2021measuring} depends on middle layers that have the smallest
effect on WikiText perplexity: on Llama and Qwen3-14B, the WikiText subsets
of \ours{} and SLEB remove them and score 24\% to 28\%, and the deep blocks of
ShortGPT keep 67.1\% and 68.4\% at 25\% removed, as in
\citet{gromov2025unreasonable} (\cref{tab:mmlu}). With an MMLU objective on
validation questions, \ours{} removes deep layers instead and is the most
accurate method at 7 of 8 settings, with 53.3\% against at most 26.7\% on
Qwen3-14B at 15 layers (\cref{app:tables}). The selection objective therefore
controls which capabilities are preserved after pruning.

\subsection{The repair Fisher score}
\label{sec:exp-score}

\begin{figure}[t]
\centering
% How well a block score ranks blocks under repair, per model: (a) mean
% within-length Spearman correlation with the exact repaired loss increase, error bars
% one standard deviation over block lengths; (b) share of block lengths where the
% lowest-scoring block is within 5% of the best block (one column).
% Data: data/score_summary.csv, written by data/export_csv.py.
\begin{tikzpicture}
\pgfplotstableread[col sep=comma,row sep=\\]{
model,model_index,predictor,predictor_index,lengths,mean_spearman,std_spearman,spearman_all,within_5pct,within_5pct_share_pct,exact_best\\
Qwen3-1.7B,1,fisher,1,6,0.5633477633477634,0.1331357581146645,0.7724581724581725,5,83.33333333333333,4\\
Qwen3-1.7B,1,cosine,2,6,0.2712842712842713,0.23415290557928073,0.802059202059202,1,16.666666666666668,1\\
Qwen3-1.7B,1,linearization,3,6,-0.5841991341991342,0.12014457224491266,-0.6635778635778636,0,0.0,0\\
Qwen3-1.7B,1,first_order,4,6,-0.5194083694083694,0.20506661505656681,-0.7976833976833978,0,0.0,0\\
gemma-3-4b-it,2,fisher,1,6,0.5271450771450772,0.37448749372464724,0.8828104744620171,5,83.33333333333333,5\\
gemma-3-4b-it,2,cosine,2,6,0.36239316239316244,0.2029051863606082,0.7859087373606432,1,16.666666666666668,0\\
gemma-3-4b-it,2,linearization,3,6,0.8627483627483628,0.041168424130498274,0.22433238268084005,6,100.0,4\\
gemma-3-4b-it,2,first_order,4,6,-0.11181318681318682,0.3377819338392054,-0.7827326938034742,0,0.0,0\\
Llama-3.1-8B-Instruct,3,fisher,1,6,0.6863442113442114,0.08304832188793697,0.9585733486943164,6,100.0,5\\
Llama-3.1-8B-Instruct,3,cosine,2,6,-0.004773004773004766,0.1750999596424542,0.7649769585253456,0,0.0,0\\
Llama-3.1-8B-Instruct,3,linearization,3,6,-0.026060051060051066,0.21165887298201314,-0.629800307219662,0,0.0,0\\
Llama-3.1-8B-Instruct,3,first_order,4,6,-0.2671134421134421,0.19247069490669352,-0.8172523041474654,0,0.0,0\\
Qwen3-14B,4,fisher,1,4,0.8586974300209594,0.06732806548445529,0.8719211822660099,3,75.0,2\\
Qwen3-14B,4,cosine,2,4,-0.20907401422107305,0.1754864046462034,0.4614337568058077,0,0.0,0\\
Qwen3-14B,4,linearization,3,4,0.0909321071085777,0.3271748801311627,-0.3006222452683433,1,25.0,0\\
Qwen3-14B,4,first_order,4,4,-0.3881532683003271,0.3782927367472914,-0.4986388384754991,0,0.0,0\\
}\scores
\begin{groupplot}[
  group style={group size=2 by 1, horizontal sep=1.15cm},
  paperstyle,
  width=0.57\columnwidth, height=0.56\columnwidth,
  xmin=0.45, xmax=4.55,
  xtick={1,2,3,4},
  xticklabels={Qwen\\1.7B, Gemma\\4B, Llama\\8B, Qwen\\14B},
  x tick label style={font=\tiny, align=center},
  major x tick style={draw=none},
  title style={font=\scriptsize, yshift=-4pt},
]
\nextgroupplot[
  title={(a) Spearman with repaired loss increase},
  ybar, bar width=0.3,
  ymin=-0.5, ymax=1.0, ytick={-0.5,0,0.5,1},
  legend to name=fisherlegend, legend columns=-1,
  legend style={/tikz/every even column/.append style={column sep=4pt}},
  legend image code/.code={\draw[#1, draw=none] (0cm,-0.08cm) rectangle (0.22cm,0.12cm);},
  extra y ticks={0}, extra y tick style={grid=major, grid style={black!50, line width=0.4pt}},
]
\addplot[bar shift=-0.17, fill=ours, draw=none,
  error bars/.cd, y dir=both, y explicit, error bar style={black!60, line width=0.4pt}, error mark options={black!60, mark size=1.2pt}]
  table[x=model_index, y=mean_spearman, y error=std_spearman,
        restrict expr to domain={\thisrow{predictor_index}}{1:1}] {\scores};
\addlegendentry{\oursfisher{} score (ours)}
\addplot[bar shift=0.17, fill=replaceme, draw=none,
  error bars/.cd, y dir=both, y explicit, error bar style={black!60, line width=0.4pt}, error mark options={black!60, mark size=1.2pt}]
  table[x=model_index, y=mean_spearman, y error=std_spearman,
        restrict expr to domain={\thisrow{predictor_index}}{2:2}] {\scores};
\addlegendentry{ReplaceMe cosine distance}
\nextgroupplot[
  title={(b) Best block found (\%)},
  ybar, bar width=0.3,
  ymin=0, ymax=112, ytick={0,25,50,75,100},
]
\addplot[bar shift=-0.17, fill=ours, draw=none,
  nodes near coords={\pgfmathprintnumber[fixed, precision=0]{\pgfplotspointmeta}},
  nodes near coords style={font=\tiny, anchor=south, inner sep=1pt}]
  table[x=model_index, y=within_5pct_share_pct,
        restrict expr to domain={\thisrow{predictor_index}}{1:1}] {\scores};
\addplot[bar shift=0.17, fill=replaceme, draw=none,
  nodes near coords={\pgfmathprintnumber[fixed, precision=0]{\pgfplotspointmeta}},
  nodes near coords style={font=\tiny, anchor=south, inner sep=1pt}]
  table[x=model_index, y=within_5pct_share_pct,
        restrict expr to domain={\thisrow{predictor_index}}{2:2}] {\scores};
\end{groupplot}
\node[anchor=south] at ($(group c1r1.north)!0.5!(group c2r1.north)+(0,0.32cm)$) {\pgfplotslegendfromname{fisherlegend}};
\end{tikzpicture}
\caption{Ranking blocks under repair without search. (a) Spearman correlation
of each score with the exact repaired loss increase of every block of a given length,
averaged over lengths (error bars: one standard deviation over lengths). (b)
Share of block lengths at which the lowest-scoring block is within 5\% of the
best block.}
\label{fig:fisher}
\end{figure}
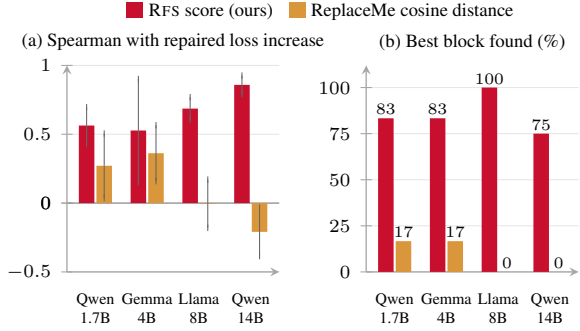

Against the exact repaired loss increase of every block of non-critical groups,
the lowest \oursfisher{} block is within 5\% of the best at 19 of 22 block
lengths; ReplaceMe's cosine distance is within 5\% at 2. The Spearman
correlation is 0.53 to 0.86 for the score and $-0.21$ to 0.36 for the cosine
distance (\cref{fig:fisher}). In \cref{tab:main} the score's block, chosen
without evaluating a pruned model, is within 3\% of \ours{} at 4 of 8 settings
and beats ReplaceMe at 7 of 8. At the largest budgets (Gemma at 13 of 34, Llama
at 12 of 32), \ours{} adds layers outside the block and reaches one half to one
third of the score's perplexity, so at heavy budgets the search justifies its
cost. The Fisher score thus provides a strong default when a single block is
sufficient, and the greedy search extends the method to heavier budgets.

\subsection{ReplaceMe's default calibration}
\label{sec:replaceme-setting}

ReplaceMe by default fits its maps on SlimOrca conversations
\citep{lian2023slimorca} without a ridge. With this fit, \ours{} has 16\% to 73\%
lower perplexity on Llama and Gemma (\cref{tab:replaceme-setting}). With a
SlimOrca objective, \ours{} differs from ReplaceMe only in which layers it
removes; its perplexity is lower at 7 of 8 settings and its accuracy at 6 of 8
(\cref{tab:slimorca}). The advantage of selection under repair therefore holds
across calibration data and fitting procedures.

\subsection{Recovery fine-tuning and a second corpus}
\label{sec:recovery}

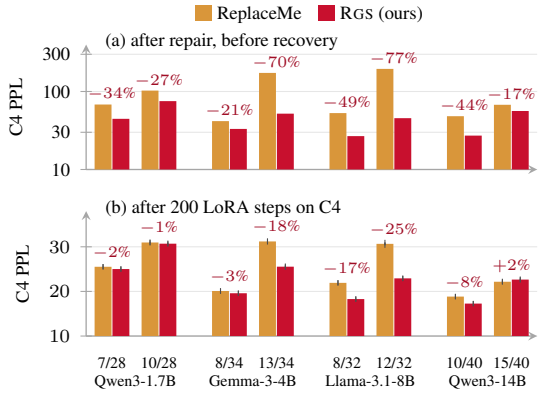
\begin{figure}[t]
\centering
% C4 perplexity of ReplaceMe and RGS (a) after repair only and (b) after 200
% LoRA steps on C4 (mean of three seeds, error bars one sample standard
% deviation), as grouped bars; the label above each RGS bar is its change
% against ReplaceMe (one column, two stacked panels).
% Data: data/recovery_c4_summary.csv, written by data/export_csv.py.
\begin{tikzpicture}
\pgfplotstableread[col sep=comma,row sep=\\]{
x,model,label,removed,depth,budget,replaceme_before_c4,replaceme_after_c4,replaceme_after_c4_std,replaceme_c4_drop,greedy_before_c4,greedy_after_c4,greedy_after_c4_std,greedy_c4_drop,greedy_before_c4_change_pct,greedy_after_c4_change_pct,replaceme_after_held_out,replaceme_after_held_out_std,greedy_after_held_out,greedy_after_held_out_std,replaceme_after_mean_without_boolq,greedy_after_mean_without_boolq\\
1.0,Qwen3-1.7B,Qwen3-1.7B,7,28,7/28,68.08145686440282,25.521736904521575,0.027964696939206866,42.55971995988125,44.628884830228486,25.013186429568865,0.045163599061777616,19.61569840065962,-34.44781165726902,-1.992617026244059,29.34725719292123,0.46123303875397426,28.769048845260897,0.6061489048743475,0.5198,0.5083333333333333\\
2.0,Qwen3-1.7B,Qwen3-1.7B,10,28,10/28,102.68566518007918,30.962791830611383,0.08149928058394977,71.7228733494678,75.24955136910422,30.696977507563076,0.0967433966555967,44.55257386154115,-26.718543199637868,-0.8584959796342106,34.44497749476526,0.3336979458743292,34.17482459434752,0.19149174105441302,0.4714,0.4727333333333333\\
3.5,gemma-3-4b-it,Gemma-3-4B,8,34,8/34,41.70217991163244,20.09522652960539,0.042065712702705864,21.60695338202705,33.12591444182008,19.58985735683721,0.01552691941662374,13.536057084982868,-20.56550877672486,-2.514871738438196,27.37250596095998,0.305384374078306,25.610066978836993,0.5582971991775533,0.5611999999999999,0.5689333333333333\\
4.5,gemma-3-4b-it,Gemma-3-4B,13,34,13/34,173.0813950495511,31.197831704852558,0.10325120364128558,141.88356334469856,51.98316145331125,25.543747017610087,0.11722448411487899,26.43941443570116,-69.96606051249523,-18.123325815502188,43.92704817028431,0.1570978046025012,35.37622036446357,0.43557234021336966,0.4847333333333334,0.5117333333333334\\
6.0,Llama-3.1-8B-Instruct,Llama-3.1-8B,8,32,8/32,52.849219534384055,21.918500403334594,0.012560594984492468,30.93071913104946,26.780715518311478,18.285479384484223,0.05347818744744285,8.495236133827255,-49.326185411522,-16.57513494079028,29.158516491908756,0.6593799454061786,21.32271574740317,0.06050553264084706,0.6272,0.5858\\
7.0,Llama-3.1-8B-Instruct,Llama-3.1-8B,12,32,12/32,194.9063440638037,30.660989716854303,0.30321821948466776,164.2453543469494,45.49219514573716,22.92624209805753,0.05262746638765795,22.565953047679635,-76.65945900106512,-25.226673014227575,39.39879819632745,0.37046962904605285,27.430392104111217,0.060126555182710144,0.5622666666666666,0.5442666666666667\\
8.5,Qwen3-14B,Qwen3-14B,10,40,10/40,48.221537528219095,18.842209557187918,0.03512157777602755,29.379327971031177,27.192273579393436,17.273508228721017,0.018835976078548367,9.91876535067242,-43.60969190689824,-8.325463761061258,19.876828392416297,0.5523679276349323,17.34462172299964,0.24309584960113223,0.5955333333333334,0.6020666666666666\\
9.5,Qwen3-14B,Qwen3-14B,15,40,15/40,67.55408204619566,22.16820508574715,0.07474418947970976,45.38587696044851,56.235533809225586,22.651642893740075,0.09832463034038122,33.58389091548551,-16.754795408558852,2.1807710914030887,22.888215589663016,0.48708886826867276,23.553711053087316,0.30378101832040794,0.5172666666666667,0.5406\\
}\recovery
\begin{groupplot}[
  group style={group size=1 by 2, vertical sep=0.55cm, xticklabels at=edge bottom},
  paperstyle,
  width=0.98\columnwidth, height=0.42\columnwidth,
  xmin=0.45, xmax=10.05,
  xtick={1,2,3.5,4.5,6,7,8.5,9.5},
  xticklabels from table={\recovery}{budget},
  x tick label style={font=\tiny},
  major x tick style={draw=none},
  title style={font=\scriptsize, at={(0.02,1)}, anchor=north west, yshift=2pt},
  label style={font=\scriptsize},
  nodes near coords style={font=\tiny, anchor=south, inner sep=1pt, text=ours!80!black},
]
\nextgroupplot[
  title={(a) after repair, before recovery},
  ybar, bar width=0.36,
  ymode=log, log origin=infty, ymin=10, ymax=400,
  ytick={10,30,100,300}, yticklabels={10,30,100,300}, yminorticks=false,
  ylabel={C4 PPL},
  legend style={at={(0.5,1.08)}, anchor=south, legend columns=-1,
    /tikz/every even column/.append style={column sep=4pt}},
  legend image code/.code={\draw[#1, draw=none] (0cm,-0.08cm) rectangle (0.22cm,0.12cm);},
]
\addplot[bar shift=-0.19, fill=replaceme, draw=none]
  table[x=x, y=replaceme_before_c4] {\recovery};
\addlegendentry{ReplaceMe}
\addplot[bar shift=0.19, fill=ours, draw=none]
  table[x=x, y=greedy_before_c4] {\recovery};
\addlegendentry{\ours{} (ours)}
% The change label, centred above the taller bar of each pair.
\addplot[bar shift=0, draw=none, fill=none, forget plot,
  nodes near coords={\pgfmathprintnumber[showpos, fixed, precision=0]{\pgfplotspointmeta}\%},
  point meta=explicit]
  table[x=x, y expr={max(\thisrow{replaceme_before_c4}, \thisrow{greedy_before_c4})},
        meta=greedy_before_c4_change_pct] {\recovery};
\nextgroupplot[
  title={(b) after 200 LoRA steps on C4},
  ybar, bar width=0.36,
  ymin=10, ymax=38, ytick={10,20,30},
  ylabel={C4 PPL},
  extra x ticks={1.5,4,6.5,9},
  extra x tick labels={Qwen3-1.7B,Gemma-3-4B,Llama-3.1-8B,Qwen3-14B},
  extra x tick style={major tick length=0pt, grid=none,
    tick label style={font=\scriptsize, yshift=-11pt}},
]
\addplot[bar shift=-0.19, fill=replaceme, draw=none,
  error bars/.cd, y dir=both, y explicit,
  error bar style={black!70, line width=0.4pt}, error mark options={black!70, mark size=1pt}]
  table[x=x, y=replaceme_after_c4, y error=replaceme_after_c4_std] {\recovery};
\addplot[bar shift=0.19, fill=ours, draw=none,
  error bars/.cd, y dir=both, y explicit,
  error bar style={black!70, line width=0.4pt}, error mark options={black!70, mark size=1pt}]
  table[x=x, y=greedy_after_c4, y error=greedy_after_c4_std] {\recovery};
% The change label, centred above the taller bar of each pair and its error bar.
\addplot[bar shift=0, draw=none, fill=none, forget plot,
  nodes near coords={\pgfmathprintnumber[showpos, fixed, precision=0]{\pgfplotspointmeta}\%},
  nodes near coords style={yshift=1.5pt},
  point meta=explicit]
  table[x=x, y expr={max(\thisrow{replaceme_after_c4}, \thisrow{greedy_after_c4})},
        meta=greedy_after_c4_change_pct] {\recovery};
\end{groupplot}
\end{tikzpicture}
\caption{C4 perplexity (a) after repair and (b) after 200 LoRA steps on C4
(mean of three seeds; error bars, one standard deviation). Labels give the
change of \ours{} against ReplaceMe: recovery shrinks the gap, and \ours{} stays
lower at 7 of 8 settings.}
\label{fig:recovery}
\end{figure}

\ours{} has lower C4 perplexity \citep{raffel2020exploring} than ReplaceMe at
both budgets on all four models, even though it selects on WikiText. After 200
LoRA steps \citep{hu2022lora} on C4 with three seeds, \ours{} keeps lower mean
perplexity at 7 of 8 settings (\cref{fig:recovery}), by 16\% to 30\% on Llama,
2.5\% to 19\% on Gemma and 1\% to 2\% on Qwen3-1.7B. The benefit of better
layer selection therefore persists after recovery fine-tuning.

\subsection{Ablations}
\label{sec:ablation}

\Cref{tab:ablation} removes one component of \ours{} at a time on the 8
settings. The repair inside the search matters most: SLEB, which searches
without it, has 19\% higher perplexity on average (up to 41\%). One-shot scoring
under repair is 21\% worse and ReplaceMe's criterion 73\% worse. Removing the
layers that \ours{} selects without the repair raises perplexity by a factor of
four, so the selected subset depends on the repair.

\begin{table}[t]
\centering
\small
\setlength{\tabcolsep}{3pt}
\resizebox{\columnwidth}{!}{%
\begin{tabular}{@{}lrrrr@{}}
\toprule
Variant & \makecell[r]{PPL /\\\ours{}} & Worst & Avg & Evals \\
\midrule
\rowcolor{oursrow} \ours{} & 1.00 & 1.00 & 48.1 & 303 \\
\quad + shift refinement & \best{0.995} & 1.00 & \best{48.4} & 303 \\
\quad $-$ repair in the search (SLEB + R) & 1.19 & 1.41 & 48.0 & 303 \\
\quad $-$ search (one-shot + R) & 1.21 & 1.73 & 46.5 & 34 \\
\quad $-$ search, block criterion (ReplaceMe) & 1.73 & 2.95 & 47.3 & 0 \\
\quad $-$ repair at deployment & 3.98 & 14.9 & -- & 303 \\
\bottomrule
\end{tabular}}
\caption{Ablation of \ours{} on the 8 settings of the four enumerated families.
PPL / \ours{}: geometric mean of the perplexity ratio to \ours{}; Worst: the
largest ratio; Avg: mean five-task accuracy (\%); Evals: mean number of pruned
models evaluated during selection.}
\label{tab:ablation}
\end{table}

\subsection{Cost}
\label{sec:cost}

The maps fold into existing weights, so at a given budget every method
in \cref{tab:main} gives a model of the same size and speed. Removing 25\% of the
layers gives $1.3\times$ speedup and removing 37.5\% gives $1.55\times$; at 24 of
64 layers, Qwen3.8-27B shrinks from 53.8 to 35.6\,GB and fits on one 48\,GB GPU
(\cref{tab:speed}). \ours{} evaluates as many pruned models as SLEB and
\oursfisher{} needs one forward and one backward pass (\cref{app:tables}).

\section{Conclusion}
\label{sec:conclusion}

The exact optimum of depth pruning depends on the repair: local
super-additivity makes a contiguous block close to the worst choice without
repair and close to the best with it. Evaluating each choice on the repaired
model that will be deployed lowers perplexity by 20\% to 66\% relative to
ReplaceMe at the cost of SLEB, and a second-order expansion of the repaired
loss gives a score computed from the intact model.

The core finding applies whenever the repair alters the cost landscape of the
removed layers. Future work can test whether it holds for low-rank or shifted
repairs. The repair Fisher score shows that repair quality can be measured from
the intact model alone, which opens a path to joint optimisation of layer
selection and the repair objective.

\section*{Limitations}

Exact optima are available only at two-layer granularity up to 6 removed groups
and at single-layer granularity up to 3; at heavier budgets we compare methods
with each other. The structural results concern perplexity: at two-layer
granularity the perplexity optimum is the most accurate subset, or within one
standard error of it, in 10 of 14 cases, and in the other four a different
subset is 2 to 7 points more accurate. We test one repair, ReplaceMe's
least-squares map; two schemes that refit later maps on the already repaired
network did not improve on it, and newer repairs
\citep{chen2025simple,chen2026prune,yun2026ghosted} are not tested. We
do not compare with BlockPruner or SubFit. Recovery used one length, 200 LoRA
steps. ReplaceMe's
example configuration fits on conversations of up to 1,024 tokens with the model
in 4 bits; we fit on blocks of 256 tokens with the model in bfloat16. We
evaluate only English text and multiple-choice tasks; multilingual text and
open-ended generation remain untested.

% Acknowledgments. ARR reviews are anonymous: keep this section out of the
% review version by leaving it empty, and fill it in for the camera-ready
% version only.

\bibliography{references}

\appendix
\section{Proof of Proposition~\ref{prop:kl}}
\label{app:proof}

Fix a position $s$ and write $p = p_s(z)$ and $q(\delta) = p_s(z+\delta)$. Then
\begin{equation*}
\mathrm{KL}(p \,\|\, q(\delta)) = \sum_y p_y \log p_y - \sum_y p_y \log q_y(\delta).
\end{equation*}
At $\delta = 0$ the divergence is zero. Its gradient is $-\sum_y p_y \nabla \log
q_y(0) = -\sum_y \nabla q_y(0) = -\nabla \sum_y q_y(0) = 0$, because $q(0) = p$
and the probabilities sum to one for every $\delta$. With $\nabla^2 \log q_y =
\nabla^2 q_y / q_y - \nabla \log q_y \nabla \log q_y^\top$ and $\sum_y \nabla^2
q_y = 0$, its Hessian is
\begin{align*}
-\sum_y p_y \nabla^2 \log q_y(0)
&= \sum_y p_y \nabla \log q_y(0)\, \nabla \log q_y(0)^\top \\
&\quad - \sum_y \nabla^2 q_y(0) \\
&= \mathbb{E}_{y \sim p}\big[\nabla \log p(y)\, \nabla \log p(y)^\top\big].
\end{align*}
Summing over positions and expanding to second order gives \cref{eq:kl}. The
remainder is $O(\|\delta\|^3)$ when the downstream network is three times
differentiable in the residual stream, which holds for transformer layers with smooth
activations and normalisation away from zero norm.

Both approximations in \cref{sec:score} are standard in the pruning literature.
The empirical Fisher replaces the model's predictions by the observed tokens;
the two agree when the model is calibrated on the calibration text. The
block-diagonal rank-one form drops the curvature between positions and, within a
position, in directions other than $g_t$. We validate the resulting score
against the exact repaired loss increases in \cref{sec:exp-score}.

\section{Algorithms}
\label{app:algorithms}

\begin{algorithm}[h]
\caption{Greedy selection under repair (\ours{})}
\label{alg:rgs}
\small
\begin{algorithmic}[1]
\Require layers $1..L$, budget $n$, objective $J$, activations of the intact model
\State $S \gets \emptyset$
\For{$k = 1, \dots, n$}
  \For{each kept layer $i \neq 1$}
    \State fit $T_{a,b}$ by \cref{eq:repair} for each run $a..b$ of $S \cup \{i\}$ \Comment{cached}
    \State $J_i \gets J(\text{model without } S \cup \{i\}\text{, repaired})$
  \EndFor
  \State $S \gets S \cup \{\argmin_i J_i\}$
\EndFor
\State \Return $S$, with its maps folded into the down-projections
\end{algorithmic}
\end{algorithm}

We always keep the first layer: the map of a run is folded into the kept layer
before the run, so a run that starts at the first layer cannot be repaired. In
every enumerated setting the first group is also critical.

\section{Mechanism of local super-additivity}
\label{app:mechanism}

\Cref{tab:mechanism} lists the explanations we tested. Only functional
similarity has partial support: given the distance, the linear CKA of the
updates of two groups also predicts their pairwise term on Qwen3-1.7B and
Llama, weakly on Qwen3-14B, and not on Gemma.

\begin{table*}[h]
\centering
\small
\begin{tabular}{p{0.24\textwidth}p{0.36\textwidth}p{0.32\textwidth}}
\toprule
Explanation & Test & Result \\
\midrule
Distribution shift of the residual stream & replace removed groups by their mean update & ruled out: pairwise terms remain ($+0.55$ against $+0.38$ for neighbours) \\
Later layers restore the update & share of the removed update that later layers restore & ruled out: Spearman $+0.17$ with the pairwise terms, against $+0.72$ for distance \\
Convexity of the loss & second-order split of each pairwise term & ruled out: curvature term $+0.24$ at long distance, where the pairwise term is zero \\
Logit effects add & exact split into loss and network parts & ruled out: joint logit changes differ from the sum by 50\% to 70\% \\
Nearby layers compute similar functions & partial Spearman of linear CKA and the pairwise term, given distance & Qwen3-1.7B $+0.37$ ($p = 0.006$), Llama $+0.36$ ($0.001$), Qwen3-14B $+0.11$ ($0.09$), Gemma $-0.06$ ($0.68$) \\
\bottomrule
\end{tabular}
\caption{Tests of explanations for local super-additivity.}
\label{tab:mechanism}
\end{table*}

\section{Search without repair}
\label{app:search}

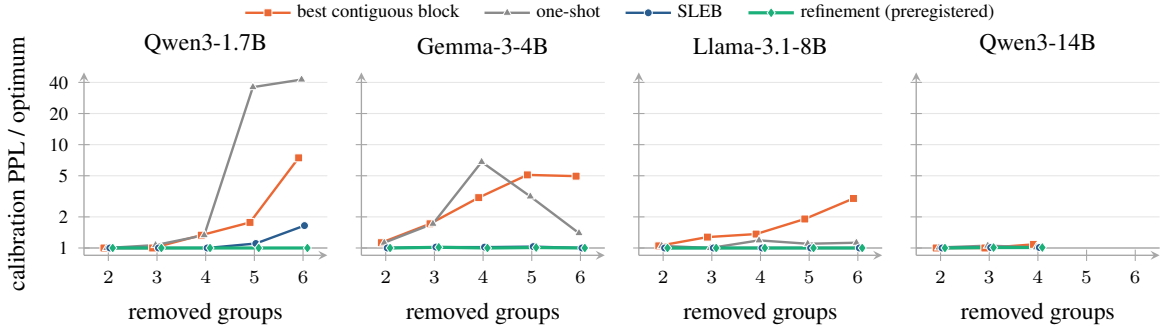
\begin{figure*}[h]
\centering
% Search without repair: calibration PPL of each method relative to the exact
% optimum, by number of removed groups, one panel per model (full width).
% Data: data/search_excess.csv, written by data/export_csv.py.
\begin{tikzpicture}
% Colours of the two search methods that are not in the main comparison.
\definecolor{bestblock}{HTML}{EB6834}
\definecolor{refine}{HTML}{1BAF7A}
\pgfplotstableread[col sep=comma,row sep=\\]{
model,model_index,removed_groups,method,method_index,excess_ratio\\
Qwen3-1.7B,1,2,block,1,1.0\\
Qwen3-1.7B,1,3,block,1,1.0\\
Qwen3-1.7B,1,4,block,1,1.3303108138789894\\
Qwen3-1.7B,1,5,block,1,1.7681085816040032\\
Qwen3-1.7B,1,6,block,1,7.450644086563554\\
Qwen3-1.7B,1,2,one_shot,2,1.0\\
Qwen3-1.7B,1,3,one_shot,2,1.066388216800626\\
Qwen3-1.7B,1,4,one_shot,2,1.3303108138789894\\
Qwen3-1.7B,1,5,one_shot,2,36.003066924283445\\
Qwen3-1.7B,1,6,one_shot,2,42.62571874555453\\
Qwen3-1.7B,1,2,sleb,3,1.0\\
Qwen3-1.7B,1,3,sleb,3,1.0\\
Qwen3-1.7B,1,4,sleb,3,1.0\\
Qwen3-1.7B,1,5,sleb,3,1.105105369478875\\
Qwen3-1.7B,1,6,sleb,3,1.6471760871149252\\
Qwen3-1.7B,1,2,refine,4,1.0\\
Qwen3-1.7B,1,3,refine,4,1.0\\
Qwen3-1.7B,1,4,refine,4,1.0\\
Qwen3-1.7B,1,5,refine,4,1.0\\
Qwen3-1.7B,1,6,refine,4,1.0\\
gemma-3-4b-it,2,2,block,1,1.125290676327302\\
gemma-3-4b-it,2,3,block,1,1.715724905185233\\
gemma-3-4b-it,2,4,block,1,3.071243921109693\\
gemma-3-4b-it,2,5,block,1,5.096778653591977\\
gemma-3-4b-it,2,6,block,1,4.955926728044638\\
gemma-3-4b-it,2,2,one_shot,2,1.125290676327302\\
gemma-3-4b-it,2,3,one_shot,2,1.715724905185233\\
gemma-3-4b-it,2,4,one_shot,2,6.805417334399246\\
gemma-3-4b-it,2,5,one_shot,2,3.1521773840860154\\
gemma-3-4b-it,2,6,one_shot,2,1.3919586299168347\\
gemma-3-4b-it,2,2,sleb,3,1.0\\
gemma-3-4b-it,2,3,sleb,3,1.0199440712392573\\
gemma-3-4b-it,2,4,sleb,3,1.0192497126616036\\
gemma-3-4b-it,2,5,sleb,3,1.0309849401465254\\
gemma-3-4b-it,2,6,sleb,3,1.0\\
gemma-3-4b-it,2,2,refine,4,1.0\\
gemma-3-4b-it,2,3,refine,4,1.015330425069403\\
gemma-3-4b-it,2,4,refine,4,1.0\\
gemma-3-4b-it,2,5,refine,4,1.009915180846888\\
gemma-3-4b-it,2,6,refine,4,1.0\\
Llama-3.1-8B-Instruct,3,2,block,1,1.0485894794830812\\
Llama-3.1-8B-Instruct,3,3,block,1,1.275032031549573\\
Llama-3.1-8B-Instruct,3,4,block,1,1.363958980048822\\
Llama-3.1-8B-Instruct,3,5,block,1,1.9082348981342716\\
Llama-3.1-8B-Instruct,3,6,block,1,3.014814461062526\\
Llama-3.1-8B-Instruct,3,2,one_shot,2,1.0485894794830812\\
Llama-3.1-8B-Instruct,3,3,one_shot,2,1.0\\
Llama-3.1-8B-Instruct,3,4,one_shot,2,1.1847423506066435\\
Llama-3.1-8B-Instruct,3,5,one_shot,2,1.099162660757142\\
Llama-3.1-8B-Instruct,3,6,one_shot,2,1.1243065824449634\\
Llama-3.1-8B-Instruct,3,2,sleb,3,1.0\\
Llama-3.1-8B-Instruct,3,3,sleb,3,1.0\\
Llama-3.1-8B-Instruct,3,4,sleb,3,1.0\\
Llama-3.1-8B-Instruct,3,5,sleb,3,1.0\\
Llama-3.1-8B-Instruct,3,6,sleb,3,1.0\\
Llama-3.1-8B-Instruct,3,2,refine,4,1.0\\
Llama-3.1-8B-Instruct,3,3,refine,4,1.0\\
Llama-3.1-8B-Instruct,3,4,refine,4,1.0\\
Llama-3.1-8B-Instruct,3,5,refine,4,1.0\\
Llama-3.1-8B-Instruct,3,6,refine,4,1.0\\
Qwen3-14B,4,2,block,1,1.0\\
Qwen3-14B,4,3,block,1,1.0\\
Qwen3-14B,4,4,block,1,1.08368281647778\\
Qwen3-14B,4,2,one_shot,2,1.0116549861681685\\
Qwen3-14B,4,3,one_shot,2,1.0531770758079988\\
Qwen3-14B,4,4,one_shot,2,1.015290785293452\\
Qwen3-14B,4,2,sleb,3,1.0\\
Qwen3-14B,4,3,sleb,3,1.0096957403741469\\
Qwen3-14B,4,4,sleb,3,1.0103443103075678\\
Qwen3-14B,4,2,refine,4,1.0\\
Qwen3-14B,4,3,refine,4,1.0096957403741469\\
Qwen3-14B,4,4,refine,4,1.0103443103075678\\
}\search
\pgfplotsset{
  % Rows of one model (model_index) and one method (method_index).
  search rows/.style 2 args={
    x filter/.expression={\thisrow{model_index}==#1 ? x : nan},
    restrict expr to domain={\thisrow{method_index}}{#2:#2},
  },
  search marks/.style={mark options={draw=white, line width=0.3pt}},
}
% Draws the four methods of panel #1.
\newcommand{\searchpanel}[1]{%
  \draw[gray!70, line width=0.5pt] (axis cs:1.5,1) -- (axis cs:6.5,1);
  \addplot[bestblock, mark=square*, mark size=1.5pt, search marks, search rows={#1}{1}]
    table[x expr=\thisrow{removed_groups}-0.09, y=excess_ratio] {\search};
  \addplot[shortgpt, mark=triangle*, mark size=1.9pt, search marks, search rows={#1}{2}]
    table[x expr=\thisrow{removed_groups}-0.03, y=excess_ratio] {\search};
  \addplot[sleb, mark=*, mark size=1.5pt, search marks, search rows={#1}{3}]
    table[x expr=\thisrow{removed_groups}+0.03, y=excess_ratio] {\search};
  \addplot[refine, mark=diamond*, mark size=2pt, line width=1.2pt, search marks,
    search rows={#1}{4}]
    table[x expr=\thisrow{removed_groups}+0.09, y=excess_ratio] {\search};
}
\begin{groupplot}[
  group style={group size=4 by 1, horizontal sep=0.45cm, yticklabels at=edge left},
  paperstyle,
  width=0.3\textwidth, height=0.25\textwidth,
  ymode=log,
  ymin=0.85, ymax=50,
  ytick={1,2,5,10,20,40},
  yticklabels={1,2,5,10,20,40},
  yminorticks=false,
  xmin=1.5, xmax=6.5,
  xtick={2,3,4,5,6},
  xlabel={removed groups},
  title style={font=\footnotesize, yshift=-3pt},
]
\nextgroupplot[title={Qwen3-1.7B}, ylabel={calibration PPL / optimum},
  legend to name=searchlegend, legend columns=-1,
  legend style={font=\scriptsize, draw=none,
    /tikz/every even column/.append style={column sep=7pt}}]
\searchpanel{1}
\legend{best contiguous block, one-shot, SLEB, refinement (preregistered)}
\nextgroupplot[title={Gemma-3-4B}]
\searchpanel{2}
\nextgroupplot[title={Llama-3.1-8B}]
\searchpanel{3}
\nextgroupplot[title={Qwen3-14B}]
\searchpanel{4}
\end{groupplot}
\node[anchor=south] at ($(group c1r1.north west)!0.5!(group c4r1.north east)+(0,0.4cm)$)
  {\pgfplotslegendfromname{searchlegend}};
\end{tikzpicture}
\caption{Calibration perplexity of each search method relative to the exact
optimum without repair, at every exactly enumerated budget; randomised methods
are averaged over 100 seeds.}
\label{fig:search}
\end{figure*}

Over the 20 exactly enumerated budgets with at least two removed groups, SLEB
finds the optimum in 13 (\cref{fig:search}). In 3 of its 7 misses its subset is
one shift of one removed group away from the optimum: the optimum keeps a group
between two removed runs that greedy removal took out early. A refinement
(iterative removal over non-critical groups followed by shift moves, at SLEB's cost),
designed on Qwen3-1.7B, finds the optimum in 16 budgets in every seed and in 2
more in 6\% and 68\% of seeds. It improves on SLEB on Gemma; on Qwen3-14B at 3 and 4 removed groups SLEB's
subset is already a local optimum under single swaps.

\section{Additional results}
\label{app:tables}

\begin{figure*}[h]
\centering
\includegraphics[width=0.8\textwidth]{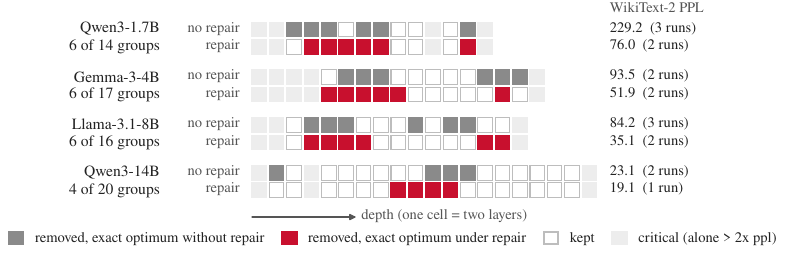}
\caption{Groups removed by the exact optimum at the heaviest enumerated budget,
without repair (grey) and with repair (red); light cells are critical groups.
Repair turns two or three separate runs into one block, or a block and a short
run, at 17\% to 67\% lower perplexity.}
\label{fig:layout}
\end{figure*}

\begin{figure*}[h]
\centering
% WikiText-2 PPL of each method relative to ReplaceMe (full width).
% Data: data/main_ppl_ratio.csv, written by data/export_csv.py.
\begin{tikzpicture}
\pgfplotstableread[col sep=comma,row sep=\\]{
x,model,label,removed,depth,budget,replaceme_ppl,shortgpt_repair_ratio,sleb_repair_ratio,fisher_block_ratio,greedy_repair_ratio,greedy_reduction_pct\\
1.0,Qwen3-1.7B,Qwen3-1.7B,7,28,7/28,53.741112509313574,1.4395802392985009,0.7504948975060775,0.7194859049563815,0.7014364837312546,29.856351626874545\\
2.0,Qwen3-1.7B,Qwen3-1.7B,10,28,10/28,68.66006362625302,1.8932234083917303,0.8608871115410518,0.7907295836815013,0.7907295836815013,20.927041631849875\\
3.5,gemma-3-4b-it,Gemma-3-4B,8,34,8/34,44.71888413017801,1.0,1.1292118295731604,0.8537918991876486,0.8034387401800515,19.656125981994855\\
4.5,gemma-3-4b-it,Gemma-3-4B,13,34,13/34,150.74365700235052,0.5786784264145267,0.458641309592945,0.6475385931189841,0.3389600889440238,66.10399110559761\\
6.0,Llama-3.1-8B-Instruct,Llama-3.1-8B,8,32,8/32,40.44330749342952,0.9733440879491746,0.5834256521504934,0.6006536463532983,0.5593755139185356,44.06244860814644\\
7.0,Llama-3.1-8B-Instruct,Llama-3.1-8B,12,32,12/32,93.97826399026906,1.0,0.4018518392799798,1.0206502029753193,0.36606165495427184,63.39383450457281\\
8.5,Qwen3-14B,Qwen3-14B,10,40,10/40,41.88116511101253,1.1515878310030763,0.6505455865224279,0.5225701684776144,0.5226195864962246,47.738041350377536\\
9.5,Qwen3-14B,Qwen3-14B,15,40,15/40,43.32041912199003,6.148727797656469,0.9943072283997434,0.7852024554102498,0.7882115426533184,21.178845734668162\\
11.0,Qwen3.8-27B,Qwen3.8-27B,16,64,16/64,28.702220311300533,1.2116558942842026,0.4597171107842072,0.4658776010006159,0.4620314571443634,53.79685428556367\\
12.0,Qwen3.8-27B,Qwen3.8-27B,24,64,24/64,36.869953839247216,1.1493850011965498,0.46491849480497,0.4271570495765241,0.43520382583365685,56.47961741663432\\
}\mainppl
\begin{axis}[
  paperstyle,
  width=\textwidth, height=0.29\textwidth,
  ymode=log, log origin y=0,
  ymin=0.21, ymax=8,
  ytick={0.3,0.5,1,2,4},
  yticklabels={0.3,0.5,1,2,4},
  yminorticks=false,
  ylabel={PPL / PPL of ReplaceMe},
  xmin=0.35, xmax=12.65,
  xtick={1,2,3.5,4.5,6,7,8.5,9.5,11,12},
  xticklabels from table={\mainppl}{budget},
  x tick label style={font=\scriptsize},
  major x tick style={draw=none},
  extra x ticks={1.5,4,6.5,9,11.5},
  extra x tick labels={Qwen3-1.7B,Gemma-3-4B,Llama-3.1-8B,Qwen3-14B,Qwen3.8-27B},
  extra x tick style={
    major tick length=0pt,
    tick label style={font=\footnotesize, yshift=-12pt},
  },
  ybar=1pt, bar width=7pt,
  legend style={at={(0.5,1.03)}, anchor=south, legend columns=-1,
    /tikz/every even column/.append style={column sep=7pt}},
  legend image code/.code={\fill[#1] (0cm,-0.08cm) rectangle (0.24cm,0.12cm);},
  clip=false,
]
\addplot[fill=shortgpt, draw=none]
  table[x=x, y=shortgpt_repair_ratio] {\mainppl};
\addlegendentry{ShortGPT + repair}
\addplot[fill=sleb, draw=none]
  table[x=x, y=sleb_repair_ratio] {\mainppl};
\addlegendentry{SLEB + repair}
\addplot[fill=ourslight, draw=ours, line width=0.4pt]
  table[x=x, y=fisher_block_ratio] {\mainppl};
\addlegendentry{\oursfisher{} (ours, score only)}
\addplot[fill=ours, draw=oursdark, line width=0.5pt,
  nodes near coords={\textminus\pgfmathprintnumber[fixed,precision=0,
    assume math mode=true]{\pgfplotspointmeta}\%},
  nodes near coords style={font=\scriptsize, color=oursdark, anchor=north,
    yshift=-0.5pt, inner sep=1pt},
  point meta=explicit]
  table[x=x, y=greedy_repair_ratio, meta=greedy_reduction_pct] {\mainppl};
\addlegendentry{\ours{} (ours)}
\addlegendimage{legend image code/.code={\draw[replaceme, dashed, line width=0.9pt] (0cm,0.02cm) -- (0.36cm,0.02cm);}}
\addlegendentry{ReplaceMe $=1$}
\draw[replaceme, dashed, line width=0.9pt]
  (axis cs:0.35,1) -- (axis cs:12.65,1);
\end{axis}
\end{tikzpicture}
\caption{Perplexity relative to ReplaceMe (log scale), every method repaired by
the same maps. \ours{} is 20\% to 66\% lower on the four enumerated families and
54\% to 56\% lower on Qwen3.8-27B.}
\label{fig:main-ppl}
\end{figure*}
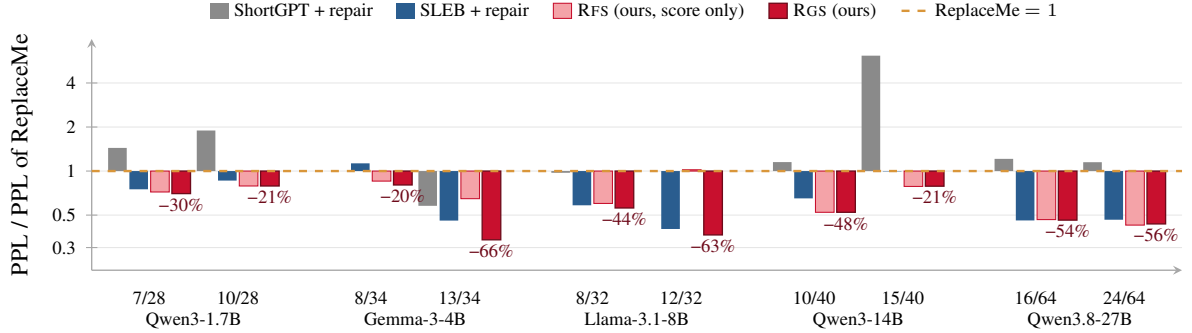

\begin{table}[h]
\centering
\small
\setlength{\tabcolsep}{4pt}
\begin{tabular}{lrrr}
\toprule
Model & $n$ & ReplaceMe & \ours{} \\
\midrule
Gemma-3-4B & 8 & 55.9 / 52.4 & \best{47.0} / \best{53.6} \\
 & 13 & 243.7 / 41.6 & \best{70.5} / \best{47.7} \\
Llama-3.1-8B & 8 & 63.7 / \best{60.2} & \best{26.1} / 56.8 \\
 & 12 & 163.1 / 48.8 & \best{44.4} / \best{51.0} \\
\bottomrule
\end{tabular}
\caption{ReplaceMe's calibration (SlimOrca conversations, unregularised least
squares) for every method. Perplexity / average accuracy (\%).}
\label{tab:replaceme-setting}
\end{table}

\begin{table}[h]
\centering
\small
\setlength{\tabcolsep}{4pt}
\begin{tabular}{lrrr}
\toprule
Model & $n$ & ReplaceMe & \ours{}, SlimOrca \\
\midrule
Qwen3-1.7B & 7 & \best{50.5} / 45.3 & 50.9 / \best{47.2} \\
 & 10 & 124.9 / \best{42.1} & \best{97.8} / 41.6 \\
Gemma-3-4B & 8 & 55.9 / 52.4 & \best{54.7} / \best{54.8} \\
 & 13 & 243.7 / 41.6 & \best{93.6} / \best{47.5} \\
Llama-3.1-8B & 8 & 63.7 / \best{60.2} & \best{28.3} / 57.7 \\
 & 12 & 163.1 / 48.8 & \best{49.3} / \best{50.7} \\
Qwen3-14B & 10 & 74.2 / 48.2 & \best{33.5} / \best{57.3} \\
 & 15 & 87.4 / 44.2 & \best{54.7} / \best{45.7} \\
\bottomrule
\end{tabular}
\caption{ReplaceMe's calibration with a SlimOrca objective for \ours{},
disjoint from the fit data; selection and fit use conversation text only.
Perplexity / average accuracy (\%).}
\label{tab:slimorca}
\end{table}

\begin{table}[h]
\centering
\small
\setlength{\tabcolsep}{3pt}
\begin{tabular}{lrrrr}
\toprule
Model & $n$ & ReplaceMe & \makecell[r]{ReplaceMe,\\task data} & \ours{}-T \\
\midrule
Qwen3-1.7B & 7 & 45.6 & 43.8 & \best{47.1} \\
 & 10 & 40.9 & 40.9 & \best{41.7} \\
Gemma-3-4B & 8 & 50.8 & 45.8 & \best{59.9} \\
 & 13 & 41.5 & 40.6 & \best{51.2} \\
Llama-3.1-8B & 8 & 57.2 & 58.9 & \best{61.2} \\
 & 12 & 46.8 & 46.8 & \best{51.9} \\
Qwen3-14B & 10 & 51.4 & 51.4 & \best{59.5} \\
 & 15 & 44.5 & 38.8 & \best{47.9} \\
\bottomrule
\end{tabular}
\caption{Average accuracy (\%) with the task objective, against ReplaceMe's
criterion computed on the hidden states of the same task training texts.}
\label{tab:task-data}
\end{table}

\begin{table}[h]
\centering
\small
\setlength{\tabcolsep}{3pt}
\begin{tabular}{lrrrr}
\toprule
Model & $n$ & ReplaceMe & \makecell[r]{\ours{}-T,\\3 tasks} & \makecell[r]{\ours{}-T,\\5 tasks} \\
\midrule
Qwen3-1.7B & 7 & 38.1 & \best{40.0} & 40.0 \\
 & 10 & 37.0 & \best{37.2} & 37.1 \\
Gemma-3-4B & 8 & 41.0 & \best{52.1} & 52.6 \\
 & 13 & 37.3 & \best{43.2} & 43.6 \\
Llama-3.1-8B & 8 & 53.4 & \best{54.3} & 54.3 \\
 & 12 & \best{47.9} & 47.3 & 47.3 \\
Qwen3-14B & 10 & 42.1 & \best{50.0} & 52.2 \\
 & 15 & 38.9 & \best{43.5} & 42.2 \\
\bottomrule
\end{tabular}
\caption{Mean accuracy (\%) on ARC-Challenge and WinoGrande when the task
objective uses only ARC-Easy, HellaSwag and PIQA (3 tasks); one standard error
is about 1.1 points. The highlight compares ReplaceMe with the 3-task
objective.}
\label{tab:unseen}
\end{table}

\begin{table}[h]
\centering
\small
\setlength{\tabcolsep}{3pt}
\begin{tabular}{lrrrr}
\toprule
Model & $n$ & C4 PPL & WikiText PPL & Avg \\
\midrule
Qwen3-1.7B & 7 & 25.5 / \best{25.0} & 29.3 / \best{28.8} & \best{52.0} / 50.8 \\
 & 10 & 31.0 / \best{30.7} & 34.4 / \best{34.2} & 47.1 / \best{47.3} \\
Gemma-3-4B & 8 & 20.1 / \best{19.6} & 27.4 / \best{25.6} & 56.1 / \best{56.9} \\
 & 13 & 31.2 / \best{25.5} & 43.9 / \best{35.4} & 48.5 / \best{51.2} \\
Llama-3.1-8B & 8 & 21.9 / \best{18.3} & 29.2 / \best{21.3} & \best{62.7} / 58.6 \\
 & 12 & 30.7 / \best{22.9} & 39.4 / \best{27.4} & \best{56.2} / 54.4 \\
Qwen3-14B & 10 & 18.8 / \best{17.3} & 19.9 / \best{17.3} & 59.6 / \best{60.2} \\
 & 15 & \best{22.2} / 22.7 & \best{22.9} / 23.6 & 51.7 / \best{54.1} \\
\bottomrule
\end{tabular}
\caption{After 200 LoRA steps on C4, mean of three seeds: ReplaceMe / \ours{}.
The spread between seeds is 0.01 to 0.3 in C4 perplexity, 0.06 to 0.66 in
WikiText perplexity and at most 0.5 points in accuracy.}
\label{tab:recovery}
\end{table}

\begin{table}[h]
\centering
\small
\setlength{\tabcolsep}{3pt}
\resizebox{\columnwidth}{!}{%
\begin{tabular}{@{}lrrrrrrr@{}}
\toprule
 & & & & \multicolumn{3}{c}{Speedup} & \\
\cmidrule(lr){5-7}
Model & Removed & Weights (GB) & KV (KB/tok.) & Prefill & Dec.\ 1 & Dec.\ 16 & PPL \\
\midrule
Qwen3-1.7B & 7/28 & 3.4 $\to$ 2.7 & 112 $\to$ 84 & 1.32 & 1.34 & 1.32 & 37.72 \\
 & 10/28 & 3.4 $\to$ 2.4 & 112 $\to$ 72 & 1.55 & 1.53 & 1.54 & 54.25 \\
Llama-3.1-8B & 8/32 & 16.1 $\to$ 12.6 & 128 $\to$ 96 & 1.32 & 1.30 & 1.31 & 22.65 \\
 & 12/32 & 16.1 $\to$ 10.8 & 128 $\to$ 80 & 1.58 & 1.54 & 1.55 & 34.43 \\
Qwen3-14B & 10/40 & 29.5 $\to$ 22.9 & 160 $\to$ 120 & 1.30 & 1.31 & 1.31 & 21.96 \\
 & 15/40 & 29.5 $\to$ 19.6 & 160 $\to$ 100 & 1.57 & 1.55 & 1.56 & 34.26 \\
Qwen3.8-27B & 16/64 & 53.8 $\to$ 41.7 & 64 $\to$ 36 & 1.29 & 1.31 & 1.29 & 13.26 \\
 & 24/64 & 53.8 $\to$ 35.6 & 64 $\to$ 32 & 1.55 & 1.58 & 1.55 & 16.03 \\
\bottomrule
\end{tabular}}
\caption{Inference cost of the folded \ours{} models in bfloat16 on RTX A6000
GPUs (two for Qwen3.8-27B): weights, KV cache per token, and speedup over the
full model for prefill of 2,048 tokens and for decoding 256 tokens at batch 1
and 16. PPL is the WikiText-2 perplexity of the folded model; it matches the
model with the maps applied as separate operations to within 0.05. Gemma-3
normalises the MLP output before the residual addition, so its maps cannot be
folded and it is not listed. Only the full-attention layers of Qwen3.8-27B keep
a KV cache.}
\label{tab:speed}
\end{table}

\begin{table}[h]
\centering
\small
\setlength{\tabcolsep}{3pt}
\resizebox{\columnwidth}{!}{%
\begin{tabular}{@{}lrrrrrrr@{}}
\toprule
 & & & & & \multicolumn{3}{c}{\ours{}, objective} \\
\cmidrule(lr){6-8}
Model (full) & Removed & ReplaceMe & ShortGPT + R & SLEB + R & WikiText & Tasks & MMLU \\
\midrule
Qwen3-1.7B (63.5) & 7 & 24.0 & \best{36.1} & 24.7 & 25.1 & 26.7 & \second{31.3} \\
 & 10 & 24.1 & 23.8 & \second{24.8} & 24.0 & 23.8 & \best{25.9} \\
Gemma-3-4B (60.6) & 8 & 24.1 & 24.1 & 27.5 & 25.4 & \second{36.3} & \best{53.0} \\
 & 13 & 23.6 & 23.8 & 23.7 & \second{25.6} & 24.7 & \best{27.2} \\
Llama-3.1-8B (68.4) & 8 & 66.3 & \second{67.1} & 28.4 & 27.0 & 48.4 & \best{67.6} \\
 & 12 & \second{58.4} & \second{58.4} & 23.5 & 24.0 & 31.6 & \best{62.3} \\
Qwen3-14B (80.8) & 10 & 23.4 & \second{68.4} & 25.9 & 24.7 & 37.5 & \best{75.1} \\
 & 15 & 25.1 & 24.6 & 26.0 & 24.0 & \second{26.7} & \best{53.3} \\
Qwen3.8-27B (83.5) & 16 & 24.1 & -- & \second{54.8} & 52.4 & \best{57.6} & -- \\
 & 24 & 24.7 & -- & \second{30.2} & 26.0 & \best{31.2} & -- \\
\bottomrule
\end{tabular}}
\caption{5-shot MMLU accuracy (\%) on 50 test questions per subject (2,850
questions) of the subsets in \cref{tab:main}, and of \ours{} with an MMLU
objective: the cross-entropy of the correct letter on 5 validation questions per
subject; the other methods select without MMLU data. Chance is 25\%. All
repairs are fitted on WikiText-2. --: not run.}
\label{tab:mmlu}
\end{table}

\paragraph{MMLU.}
\Cref{tab:mmlu} gives 5-shot MMLU accuracy for every subset. The MMLU objective
uses 5 questions per subject from the validation split, 285 in total; the test
questions and the dev split, which supplies the 5-shot examples, are disjoint
from them. With it \ours{} is the most accurate method at 7 of 8 settings, by
0.5 to 26.6 points. The exception is Qwen3-1.7B at 7 layers, where ShortGPT
reaches 36.1\% and \ours{} 31.3\%. On Qwen3-1.7B at 10 and Gemma at 13 layers
all methods are within 3 points of chance. Each objective favours the metric that it optimises: the MMLU subsets
have 1.7 to 4.9 times the WikiText perplexity of the WikiText subsets, and on
Qwen3-14B a five-task accuracy 3 to 10 points lower. At 25\% removed, the
WikiText subsets of \ours{} and SLEB on Llama and Qwen3-14B put 0.80 to 0.93 of
the probability on the four answer letters, close to the full model's 0.999,
and each answers its most frequent letter for at most 56\% of the questions, so
they follow the answer format. ReplaceMe's block on Qwen3-14B at 10 layers puts
0.06 on the letters and answers A for 80\% of the questions. On
Qwen3.8-27B the task objective gives the highest MMLU accuracy at both budgets:
57.6\% at 16 layers against 54.8\% for SLEB and 24.1\% for ReplaceMe, and
31.2\% at 24 layers against 30.2\% for SLEB. The gain of the task objective thus
carries over from the zero-shot tasks to knowledge questions on the largest
model.

\paragraph{Selection time.}
On Llama-3.1-8B at 12 of 32 layers on one RTX A6000, the selection takes 384
seconds for \ours{} (319 pruned models), 343 for SLEB with repair, 487 for
LaCo, 56 for Prune\&Comp, 33 for \oursfisher{}, 8 for ReplaceMe and 1 for
ShortGPT. Each repaired method also needs 29 seconds to collect the activations
for the maps.

\paragraph{Calibration size.}
With 16 calibration sequences the optimum on the calibration sequences is also
the optimum on the test sequences in 13 of 17 enumerated settings. Resampling
calibration sets from a pool of 64 shows that the disagreement is sampling
noise: agreement rises to 100\% at 64
sequences on Qwen3-1.7B and Gemma-3-4B.

\paragraph{Repair fitting.}
Two schemes that fit later maps on the already repaired network did not improve
on ReplaceMe's map. Fitting towards the activations of the intact network made
multi-run subsets 1\% to 16\% worse on Qwen3-1.7B and Llama; fitting towards the
output of the removed run, computed on the inputs of the repaired network, was
about 5\% better on Qwen3-1.7B and within 1.5\% either way on Llama. These
differences are small compared with the differences between selection methods
in \cref{tab:main}, so the gains of \ours{} come from the choice of layers.

\section{Per-task results}
\label{app:per-task}

\begin{table*}[t!]
\centering
\setlength{\tabcolsep}{3.6pt}
\renewcommand{\arraystretch}{0.98}
\resizebox{\textwidth}{!}{%
\begin{tabular}{l rrrrrrr c rrrrrrr}
\toprule
 & \multicolumn{7}{c}{25\% of layers removed} & & \multicolumn{7}{c}{37.5\% of layers removed} \\
\cmidrule(lr){2-8}\cmidrule(l){10-16}
Method & PPL$\downarrow$ & ARC-e & ARC-c & HS & PIQA & WG & Avg$\uparrow$
 & & PPL$\downarrow$ & ARC-e & ARC-c & HS & PIQA & WG & Avg$\uparrow$ \\
\midrule
\multicolumn{16}{l}{\textit{Qwen3-1.7B, 28 layers: 7 and 10 removed}} \\
Full model & 26.5 & 70.0 & 42.6 & 53.9 & 72.8 & 60.7 & 60.0 & & 26.5 & 70.0 & 42.6 & 53.9 & 72.8 & 60.7 & 60.0 \\
ShortGPT + R & 77.4 & 46.6 & \best{29.8} & 38.4 & 61.0 & \best{54.6} & \second{46.1} & & 130.0 & 35.8 & \best{24.4} & 34.6 & 56.0 & 51.2 & 40.4 \\
ReplaceMe & 53.7 & \second{47.1} & 25.5 & 42.4 & 62.6 & 50.6 & 45.6 & & 68.7 & 34.4 & 22.4 & 38.2 & \second{57.6} & 51.7 & 40.9 \\
SLEB + R & 40.3 & 42.8 & 26.0 & \best{43.5} & \best{64.3} & 52.7 & 45.9 & & \second{59.1} & \second{36.2} & 22.0 & \best{40.0} & \best{58.2} & \second{52.5} & \best{41.8} \\
\rowcolor{oursrow} \oursfisher{} (ours) & \second{38.7} & 42.4 & 24.3 & \second{42.7} & 62.0 & \second{53.0} & 44.9 & & \best{54.3} & 35.8 & 21.4 & \second{38.6} & 56.7 & \best{53.0} & 41.1 \\
\rowcolor{oursrow} \ours{} (ours) & \best{37.7} & 41.6 & 24.1 & 41.8 & 61.7 & 51.7 & 44.2 & & \best{54.3} & 35.8 & 21.4 & \second{38.6} & 56.7 & \best{53.0} & 41.1 \\
\rowcolor{oursrow} \ours{}-T (ours) & 64.5 & \best{49.3} & \second{28.0} & \second{42.7} & \second{63.5} & 52.0 & \best{47.1} & & 110.0 & \best{39.9} & \second{23.6} & 37.6 & 56.8 & 50.7 & \second{41.7} \\
\midrule
\multicolumn{16}{l}{\textit{Gemma-3-4B, 34 layers: 8 and 13 removed}} \\
Full model & 32.0 & 78.9 & 57.1 & 65.2 & 78.1 & 69.6 & 69.8 & & 32.0 & 78.9 & 57.1 & 65.2 & 78.1 & 69.6 & 69.8 \\
ShortGPT + R & 44.7 & 53.6 & 24.6 & 52.2 & 66.3 & \second{57.5} & 50.8 & & 87.2 & 40.2 & 22.9 & 41.8 & 58.1 & 50.4 & 42.7 \\
ReplaceMe & 44.7 & 53.6 & 24.6 & 52.2 & 66.3 & \second{57.5} & 50.8 & & 150.7 & 36.8 & 23.3 & 40.4 & 55.6 & 51.3 & 41.5 \\
SLEB + R & 50.5 & \second{61.4} & \second{32.1} & \second{52.8} & \best{70.6} & 56.7 & \second{54.7} & & \second{69.1} & 42.3 & \second{25.4} & 42.8 & \second{61.5} & 51.7 & \second{44.7} \\
\rowcolor{oursrow} \oursfisher{} (ours) & \second{38.2} & 46.8 & 24.1 & 49.0 & 66.9 & 53.6 & 48.1 & & 97.6 & 34.9 & 20.5 & 38.4 & 56.1 & \best{53.1} & 40.6 \\
\rowcolor{oursrow} \ours{} (ours) & \best{35.9} & 52.3 & 27.1 & 51.8 & 69.0 & 53.0 & 50.6 & & \best{51.1} & \second{42.7} & 22.5 & \second{44.3} & \second{61.5} & \second{52.3} & \second{44.7} \\
\rowcolor{oursrow} \ours{}-T (ours) & 68.1 & \best{69.5} & \best{44.2} & \best{55.1} & \second{69.9} & \best{61.0} & \best{59.9} & & 143.7 & \best{59.4} & \best{34.1} & \best{44.4} & \best{65.2} & \best{53.1} & \best{51.2} \\
\midrule
\multicolumn{16}{l}{\textit{Llama-3.1-8B-Instruct, 32 layers: 8 and 12 removed}} \\
Full model & 13.7 & 79.7 & 55.6 & 71.0 & 82.5 & 73.2 & 72.4 & & 13.7 & 79.7 & 55.6 & 71.0 & 82.5 & 73.2 & 72.4 \\
ShortGPT + R & 39.4 & 59.3 & \second{40.3} & 56.5 & 70.4 & \best{71.1} & \second{59.5} & & 94.0 & 38.2 & \second{31.6} & 42.0 & 58.1 & \best{64.1} & 46.8 \\
ReplaceMe & 40.4 & 56.3 & 37.2 & 54.4 & 68.2 & \second{69.7} & 57.2 & & 94.0 & 38.2 & \second{31.6} & 42.0 & 58.1 & \best{64.1} & 46.8 \\
SLEB + R & \second{23.6} & \second{61.3} & 34.0 & \best{58.7} & 72.6 & 54.5 & 56.2 & & \second{37.8} & 46.6 & 26.3 & \second{48.5} & \second{66.0} & 54.6 & 48.4 \\
\rowcolor{oursrow} \oursfisher{} (ours) & 24.3 & 52.4 & 27.8 & 54.3 & 69.4 & 51.9 & 51.2 & & 95.9 & 32.7 & 23.7 & 37.1 & 54.0 & 49.7 & 39.4 \\
\rowcolor{oursrow} \ours{} (ours) & \best{22.6} & 59.9 & 31.3 & \second{57.7} & \best{73.5} & 54.4 & 55.4 & & \best{34.4} & \second{48.4} & 25.6 & \best{48.7} & \best{67.1} & 55.2 & \second{49.0} \\
\rowcolor{oursrow} \ours{}-T (ours) & 34.6 & \best{68.1} & \best{41.1} & 56.1 & \second{73.1} & 67.6 & \best{61.2} & & 54.0 & \best{52.3} & \best{32.0} & 47.5 & 65.3 & \second{62.6} & \best{51.9} \\
\midrule
\multicolumn{16}{l}{\textit{Qwen3-14B, 40 layers: 10 and 15 removed}} \\
Full model & 15.0 & 83.2 & 60.8 & 67.9 & 80.3 & 73.0 & 73.0 & & 15.0 & 83.2 & 60.8 & 67.9 & 80.3 & 73.0 & 73.0 \\
ShortGPT + R & 48.2 & 46.8 & 32.0 & 46.0 & 60.2 & \best{61.1} & 49.2 & & 266.4 & 30.1 & 23.6 & 30.5 & 50.7 & \second{53.7} & 37.7 \\
ReplaceMe & 41.9 & 55.0 & 30.6 & 51.3 & 66.6 & 53.7 & 51.4 & & 43.3 & 42.5 & \second{26.6} & 43.1 & 59.1 & 51.2 & 44.5 \\
SLEB + R & \second{27.2} & 48.6 & 28.1 & 49.7 & 65.7 & 53.6 & 49.1 & & 43.1 & 37.9 & 24.2 & 43.8 & 57.4 & 51.7 & 43.0 \\
\rowcolor{oursrow} \oursfisher{} (ours) & \best{21.9} & 54.8 & \second{32.1} & 52.4 & \best{71.4} & 54.2 & 53.0 & & \best{34.0} & \second{46.3} & 24.8 & \second{44.6} & \second{62.0} & 51.4 & \second{45.8} \\
\rowcolor{oursrow} \ours{} (ours) & \best{21.9} & \second{62.9} & 31.5 & \best{54.1} & 71.0 & 52.5 & \second{54.4} & & \second{34.1} & 43.2 & 23.9 & \best{46.8} & 60.2 & 51.6 & 45.1 \\
\rowcolor{oursrow} \ours{}-T (ours) & 37.9 & \best{68.5} & \best{44.4} & \second{53.4} & \second{71.2} & \second{60.1} & \best{59.5} & & 67.7 & \best{49.2} & \best{30.0} & 43.5 & \best{62.4} & \best{54.5} & \best{47.9} \\
\midrule
\multicolumn{16}{l}{\textit{Qwen3.8-27B, hybrid attention, 64 layers: 16 and 24 removed}} \\
Full model & 10.1 & 74.4 & 58.5 & 74.1 & 81.9 & 75.3 & 72.8 & & 10.1 & 74.4 & 58.5 & 74.1 & 81.9 & 75.3 & 72.8 \\
ShortGPT + R & 34.8 & 33.6 & 23.1 & 41.3 & 53.3 & 51.5 & 40.6 & & 42.4 & 32.3 & 21.8 & 38.1 & 52.4 & 48.5 & 38.6 \\
ReplaceMe & 28.7 & 33.8 & 22.0 & 42.5 & 58.3 & 54.4 & 42.2 & & 36.9 & 36.4 & 24.5 & 42.5 & 55.7 & 51.9 & 42.2 \\
SLEB + R & \best{13.2} & \second{74.8} & \second{49.6} & 62.6 & \best{77.9} & \second{63.8} & \second{65.7} & & 17.1 & \second{65.8} & \best{37.7} & \best{56.1} & \second{71.3} & \second{56.0} & \second{57.4} \\
\rowcolor{oursrow} \oursfisher{} (ours) & 13.4 & \best{77.4} & 48.4 & \best{63.3} & \second{77.4} & \best{65.3} & \best{66.4} & & \best{15.7} & 58.1 & 34.1 & 55.3 & 71.2 & \best{58.8} & 55.5 \\
\rowcolor{oursrow} \ours{} (ours) & \second{13.3} & 73.8 & \best{49.7} & \second{63.0} & 77.3 & 60.6 & 64.9 & & \second{16.0} & \best{66.9} & \second{37.2} & \second{55.6} & \best{74.2} & 55.5 & \best{57.9} \\
\bottomrule
\end{tabular}}
\caption{Per-task results behind \cref{tab:main}: WikiText-2 perplexity (PPL)
and zero-shot accuracy (\%) on ARC-Easy, ARC-Challenge, HellaSwag (HS), PIQA
and WinoGrande (WG), 1,000 items each. Within each model, the best pruned
result is highlighted and the second best underlined.}
\label{tab:per-task}
\end{table*}

Per-task accuracy shows where the WikiText objective and task accuracy disagree.
At 25\% removed, ShortGPT's and ReplaceMe's deep blocks keep WinoGrande on Llama
(71.1 and 69.7 against 54.4 for \ours{}), while \ours{} with the WikiText
objective removes middle layers that WinoGrande needs but that have little
effect on perplexity.
The task objective recovers this: \ours{}-T reaches 67.6 on WinoGrande there,
has the best ARC-Easy accuracy on all four families at both budgets, and the best
ARC-Challenge accuracy at 6 of these 8 settings.

\section{Implementation details}
\label{app:details}

\begin{table}[h]
\centering
\small
\setlength{\tabcolsep}{4pt}
\begin{tabular}{lrrl}
\toprule
Model & Layers & Groups & Subsets per budget \\
\midrule
Qwen3-1.7B & 28 & 14 & all, up to 6 removed \\
Qwen3-1.7B & 28 & 28 & all, up to 3 removed \\
Gemma-3-4B & 34 & 17 & all, up to 6 removed \\
Llama-3.1-8B & 32 & 16 & all, up to 6 removed \\
Qwen3-14B & 40 & 20 & all, up to 4 removed \\
\bottomrule
\end{tabular}
\caption{Exactly enumerated settings: every subset of groups up to the stated
number of removed groups is evaluated on every calibration and test sequence,
without and with repair; about 53,000 subsets in total.}
\label{tab:enumeration}
\end{table}

Models run in bfloat16. Sequences begin with a BOS token when the tokenizer adds
one. Repair maps are solved in float32 with the ridge; without a ridge they are
solved in float64 with a pseudo-inverse, because Gemma's MLP outputs on SlimOrca
are rank-deficient. For Gemma~3 the map acts on the output of the
post-feedforward normalisation, which cannot be folded into a weight, so the
repaired Gemma model has one extra $d \times d$ matrix per run. Qwen3.8-27B runs
split over two GPUs. During the study we found and fixed two faults, each
recorded with its effect in the repository: removed or moved Gemma layers
received the wrong attention type, and unregularised maps were first solved by a
routine that assumes full rank. Every number in this paper comes from runs after
the fixes. All data behind the tables and figures is exported to CSV files in
the supplementary material.

\end{document}